\documentclass[journal, onecolumn, draftcls]{IEEEtran}
\usepackage[english]{babel}

\usepackage{cite}
\usepackage{epstopdf}

\ifCLASSINFOpdf
\else
\fi
\usepackage[T1]{fontenc}
\usepackage[latin9]{inputenc}
\usepackage{amsmath}
\usepackage{amsthm}
\usepackage{amssymb}
\usepackage{stmaryrd}
\usepackage{graphicx}
\usepackage{color,xcolor}
\newtheorem{thm}{Theorem}
\newtheorem{problem}{Problem}
\newtheorem{fact}{Assumption}

\newtheorem{rem}{Remark}
\newtheorem{lem}{Lemma}
\newtheorem{prop}{Proposition}

\begin{document}
	
	\title{Semi-global Periodic Event-triggered Output Regulation for Nonlinear
		Multi-agent Systems }
	
	\author{Shiqi Zheng, \textit{Member IEEE}, Peng Shi, \textit{Fellow IEEE}, Huiyan Zhang
\thanks{The work was supported by the National Natural Science Foundation
	of China (Grant No. 61703376). }
\thanks{S. Zheng is with the School of Automation, China University of Geosciences, and
	Hubei key Laboratory of Advanced Control and Intelligent Automation for Complex Systems, Wuhan, China (e-mail: zhengshiqi1000@foxmail.com).}
\thanks{P. Shi is with the School of Electrical and Electronic Engineering, The University of Adelaide, Adelaide, SA 5005, Australia, and also with the
	College of Engineering and Science, Victoria University, Melbourne VIC 8001, Australia (e-mail: peng.shi@adelaide.edu.au).}
\thanks{H. Zhang is with the National Research Base of Intelligent Manufacturing Service, Chongqing Technology and Business University, Chongqing, 400067, China (e-mail: huiyanzhang@ctbu.edu.cn).}

}
	\maketitle
	\begin{abstract}
		This study focuses on periodic event-triggered (PET) cooperative output
		regulation problem for a class of nonlinear multi-agent systems. The
		key feature of PET mechanism is that event-triggered conditions are
		required to be monitored only periodically. This approach is beneficial
		for Zeno behavior exclusion and saving of battery energy of onboard
		sensors. At first, new PET distributed observers are proposed to estimate
		the leader information. We show that the estimation error converges
		to zero exponentially with a known convergence rate under asynchronous
		PET communication. Second, a novel PET output feedback controller
		is designed for the underlying strict feedback nonlinear multi-agent
		systems. Based on a state transformation technique and a local PET
		state observer, the cooperative semi-global output regulation problem
		can be solved by the proposed new control design technique. Simulation
		results of multiple Lorenz systems illustrate that the developed control
		scheme is effective. 
	\end{abstract}
	

	\section{Introduction}
	
	Output regulation problem has attracted an increasing attention recently.
	Output regulation aims to make tracking error converge to zero while
	rejecting disturbance. Reference and disturbance signals are produced
	by an exosystem. For typical examples, the internal model principle
	was used for the output regulation of linear multi-variable systems
	\cite{key-1}. \cite{key-2,key-3,key-3b} focused on the output regulation
	problem of nonlinear systems. Different classes of nonlinear systems,
	such as first order system, output feedback system and strict feedback
	system, were considered.
	
	The output regulation theory has also been shown to be a powerful
	method for multi-agent systems \cite{key-3a,key-3aa,key-3aaa,key-ad,key-ad1}.
	On the basis of this theory, the leader-following problem can be handled
	effectively despite parametric uncertainties and external disturbance.
	For instance, in \cite{key-4,key-5}, the cooperative output regulation
	problem for linear and nonlinear multi-agent systems were solved using
	distributed observer technique.
	
	With the continuous development of embedded microprocessors in engineering
	system, a critical issue for multi-agent systems is reducing the communication
	burden. Apparently, continuous communication may be unrealistic in
	most applications because the bandwidth and energy are limited. Event-triggered
	control strategy has been lately introduced for the cooperative control
	of multi-agent systems \cite{key-6-1,key-666}. The idea of the event-triggered
	control is that data transmission is conducted only under some certain
	conditions. Event-triggered mechanism is an effective method for resource-limited
	applications. A number of works on various kinds of event-triggered
	control methods \cite{key-5a,key-5aa,key-5aa-1} have been conducted.
	
	More recently, in \cite{key-6,key-7}, a new periodic event-triggered
	(PET) control method has been presented. Different from other event-triggered
	mechanisms, PET mechanism is required to monitor data communication
	and triggered conditions only at discrete sampling instants. This
	characteristic brings some promising advantages (see \cite{key-11a}).
	First, the inter-event time naturally becomes multiples of sampling
	periods. This condition not only \textit{strictly} excludes the Zeno
	behavior but is also useful for digital implementation where tasks
	are always executed periodically. Second, the energy for evaluating
	the event-triggered condition can also be saved given that no continuous
	monitoring exists. This condition is beneficial for saving the battery
	energy of onboard sensors. \textit{However, to the best of our knowledge,
		the PET cooperative output regulation problem for nonlinear multi-agent
		systems has not been fully investigated. }
	
	Inspired by the above observation, in this paper we investigate the
	problem of PET cooperative output regulation for a class of nonlinear
	multi-agent systems. The main challenges are as follows:
	
	1) The communication of multi-agent systems is assumed to be asynchronous.
	That is, each agent may have different sampling times and transmit
	data asynchronously. Thus, the existing distributed observers \cite{key-5,key-8,key-9}
	become invalid;
	
	2) Each agent is described by a high order strict feedback nonlinear
	system. Moreover, only the output information of each agent is available.
	This setup is more general than the existing works \cite{key-5,key-10,key-11}
	(see Remark \ref{rem:Compared-with-the}); and
	
	3) Note that the sampled data control can be regarded as a special
	case of the PET control. However, very few works have been conducted
	on sampled data output regulation for nonlinear systems, not to mention
	the PET control. In fact, only recently, the PET/sampled data output
	regulation problem has been solved for linear systems \cite{key-6aaa,key-11a}.
	The nonlinear dynamics of the considered systems will cause many difficulties
	to the PET output regulation problem.
	
	To overcome the these difficulties, we provide our main contributions
	as follows:
	\begin{itemize}
		\item New PET distributed observers are proposed to estimate the leader
		information. On the basis of the properties of time-delay systems,
		exponential functions and matrix norms, we demonstrate that the estimation
		error will converge to zero exponentially with a known convergence
		rate under asynchronous PET communication.
		\item A novel PET output feedback controller is presented for the strict
		feedback nonlinear multi-agent systems. Based on a state transformation
		technique and a local PET observer, we show that the proposed PET
		output feedback controller can solve the cooperative semi-global output
		regulation problem. Lyapunov function in logarithm form and Gronwall's
		inequality are skillfully used to prove this result.
	\end{itemize}

	This paper is organized as follows. Section II presents problem formulation
	and preliminaries. New PET distributed observer and PET control law
	are provided in Sections III and IV respectively. In Section V simulation
	results of multiple Lorenz systems are presented to demonstrate the
	effectiveness of the proposed new design scheme.  The conclusion is
	drawn in Section VI. Detailed proofs are put in the Appendices.
	
	\textit{Notations.} For a matrix $X_{i}\in\mathbb{R}^{n_{i}\times m}(i=1,2,...,N)$,
	$\mathrm{col}(X_{1},X_{2},...,X_{N})=[X_{1}^{\mathrm{T}}\thinspace X_{2}^{\mathrm{T}}\thinspace...X_{N}^{\mathrm{T}}]^{\mathrm{T}}$.
	For $A\in\mathbb{R}^{n\times m}$, $\mathrm{vec}(A)=\mathrm{col}(a_{1},...,a_{n})$
	where $a_{i}\in\mathbb{R}^{n\times1}$ is the $i$th column of $A$.
	$\mathrm{sat_{\mathit{R}}}(x):\mathbb{R}\rightarrow\mathbb{R}$ with
	a positive constant $R$ represents the saturation function, that
	is $\mathrm{sat_{\mathit{R}}}(x)=x$ if $|x|\leq R$, $\mathrm{sat_{\mathit{R}}}(x)=R$
	if $x>R$ and $\mathrm{sat_{\mathit{R}}}(x)=-R$ if $x<-R$. Given
	a time-varying matrix $B(t)\in\mathbb{R}^{n\times m}$, \textcolor{black}{define
		a set $\mathsf{E}(\gamma)$ with a positive constant $\gamma$.} If
	$B(t)\in\mathsf{E}(\gamma)$ then $||B(t)||$ converges to zero exponentially,
	that is, $||B||\leq c\mathrm{e}^{-\gamma t}$ for $\forall t\in[0,+\infty)$
	where $c$ is a positive constant.
	
	\section{Problem formulation and preliminaries}
	
	\subsection{Problem formulation}
	
	The following multi-agent systems consisting of one leader and $N$
	followers are considered. The leader is expressed as:
	\begin{align}
	\dot{\nu} & =A\nu,\label{eq:1}\\
	{\color{black}y_{0}} & {\color{black}=q_{0}(\nu)}
	\end{align}
	where $\nu\in\mathbb{R}^{n_{\nu}}$ is the reference signal and/or
	external disturbance with $n_{\nu}\in\mathbb{N}$. $y_{0}\in\mathbb{R}$
	is the output of the leader. $A$ is a given system matrix, $q_{0}(\nu)$
	is a sufficiently smooth function with ${\color{black}q_{0}(0)=0}$.
	Meanwhile, assume that there exists a known compact set $\mathbb{V}\subseteq\mathbb{R}^{n_{\nu}}$
	such that $\nu\in\mathbb{V}$.
	
	The followers are given by strict feedback nonlinear systems:
	\begin{align}
	\dot{z}_{i} & =f_{i0}(z_{i},x_{i1},\nu,w),\nonumber \\
	\dot{x}_{ij} & =f_{ij}(z_{i},x_{i1},...,x_{ij},\nu,w)+b_{ij}(w)x_{i,j+1},\nonumber \\
	\dot{x}_{in} & =f_{in}(z_{i},x_{i1},...,x_{in},\nu,w)+b_{in}(w)u_{i},\label{eq:3}\\
	y_{i} & =x_{i1},\thinspace j=1,2,...,n-1\nonumber 
	\end{align}
	where $i\in\{1,2,...,N\}$. $n\in\mathbb{N}$ is the order of the
	$i$th subsystem, $z_{i}\in\mathbb{R}^{n_{z_{i}}}$ and $x_{ij},x_{in}\in\mathbb{R}$
	denote the system states with $n_{z_{i}}\in\mathbb{N}$, \textcolor{black}{$y_{i}\in\mathbb{R}$
		is the system output.} $w\in\mathbb{R}^{n_{w}}$ represents uncertain
	parameters with $n_{w}\in\mathbb{N}$. Also assume that there exists
	a known compact set $\mathbb{W}\subseteq\mathbb{R}^{n_{w}}$ such
	that $w\in\mathbb{W}$. $f_{i0}(\cdot),f_{ij}(\cdot),b_{ij}(w)(i=1,...,N;j=1,...,n)$
	are sufficiently smooth nonlinear functions with $f_{i0}(0,...,0,w)=0,f_{ij}(0,...,0,w)=0$
	and $b_{ij}(w)>0$ for $\forall w\in\mathbb{W}$. 
	
	A directed graph $\mathcal{G}$ is used to describe the communication
	for the multi-agent systems. Let $\mathcal{G}=(\mathcal{V},\mathcal{E})$
	where $\mathcal{V}=\{1,2,...,N\}$ denotes the set of vertices and
	$\mathcal{E}\subseteq\mathcal{V}\times\mathcal{V}$ represents the
	set of edges. Matrix $\widetilde{A}=[a_{ij}]\in\mathbb{R}^{N\times N}$
	is defined, such that if $(j,i)\in\mathcal{E}$ then $a_{ij}=1$,
	otherwise $a_{ij}=0$. Laplacian matrix is defined as $\mathcal{L}=\widetilde{D}-\widetilde{A}$
	with $\widetilde{D}=\mathrm{diag}(\tilde{d}_{1},\tilde{d}_{2},...,\tilde{d}_{N})$
	and $\tilde{d}_{i}=\sum_{j=1}^{N}a_{ij}(i\in\mathcal{V})$. For communication
	between the leader and followers, $a_{i0}$ is defined such that if
	the followers can have access to the leader, then $a_{i0}=1$; otherwise
	$a_{i0}=0$. This indicates that only a small number of followers
	can obtain the information of the leader. Finally, we assume that
	there exists a directed spanning tree for the considered graph with
	the leader as the root. Then, we know $-\mathcal{H}=-(\mathcal{L}+\widetilde{B})$
	is Hurwitz with $\widetilde{B}=\mathrm{diag}(a_{10},a_{20},...,a_{N0}).$
	
	The problem we are going to solve is formulated as follows:
	\begin{problem}
		\label{prob:Given-a-multi-agent}\textcolor{black}{(}\textit{\textcolor{black}{Cooperative
				semi-global output regulation problem}}\textcolor{black}{)} \textcolor{black}{Consider
			the multi-agent systems (\ref{eq:1})-(\ref{eq:3}) with their corresponding
			graph $\mathcal{G}$. Suppose that the initial states $\nu(0),z_{i}(0),x_{ij}(0)$
			of the  systems belong to a given compact set. The control
			objective is to design a PET distributed output feedback control law
			for each follower such that }
		
		\textcolor{black}{1) All the signals are uniformly bounded for $\forall t\in[0,+\infty)$;
			and,}
		
		\textcolor{black}{2) The output regulation error $e_{i}(t)\triangleq y_{i}(t)-y_{0}(t)$
			converges to zero exponentially, $i.e.$, $\underset{t\rightarrow+\infty}{\lim}|e_{i}(t)|\rightarrow0$
			$\forall i\in\{1,2,...,N\}$.}
	\end{problem}
	\begin{rem}
		\textcolor{black}{The signals in Problem 1 are all the signals
			in the closed-loop control system. They include all the states $x_{ij}(i=1,...,n;j=1,...N)$
			of the followers, the control input $u_{i}$, the variables $\hat{\nu}_{i},\hat{\xi}_{i}$
			in the proposed distributed observer and state observer in Sections
			III-IV $etc$.}
	\end{rem}
	\begin{rem}
		\label{rem:Compared-with-the}\textcolor{black}{Contrary to the existing
			works, the considered problem is more general and practical. The reasons
			are as follows:}
		
		\textcolor{black}{1) System (\ref{eq:3}) is in a high order strict
			feedback form. The strict feedback nonlinear systems is more general
			than many other kinds of nonlinear systems \cite{key-5,key-8,key-11,key-11aa},
			such as linear systems, low order nonlinear systems, normal form nonlinear
			systems $etc$.}
		
		\textcolor{black}{2) Different from \cite{key-9,key-10} that deals
			with state feedback control, we consider output feedback control problem
			for nonlinear systems. The problem becomes more involved because only
			output information of the high order nonlinear system (\ref{eq:3})
			is available. }
		
		\textcolor{black}{3) The PET data transmission is considered. This
			means that only PET output information is available for the controller
			design, which will further complicate the design and analysis process.}
		
		\textcolor{black}{4) The output regulation problem is examined, where
			reference tracking, disturbance rejection and parametric uncertainties
			are simultaneously considered. Our study extends the results in \cite{key-21,key-211},
			where only reference tracking problem is studied.}
	\end{rem}
	
	\subsection{Preliminaries}
	
	We introduce some basic assumptions and useful results.
	
	\textit{1) Leader}
	
	For the leader dynamic of (\ref{eq:1}), assume
	\begin{fact}
		\label{fact:4} $A$ in (\ref{eq:1}) is a skew-symmetric matrix whose
		eigenvalues are semi-simple with zero real parts.
	\end{fact}
	\begin{rem}
		Assumption 1 is standard in output regulation problem \cite{key-3,key-9}.
		When $A$ is neutrally stable, a large class of commonly used signals
		$\nu$, such as sine, cosine and constant signals, can be produced.
	\end{rem}
	\textit{2) Followers}
	
	For the nonlinear system (\ref{eq:3}), we have:
	\begin{fact}
		\label{fact:5} $f_{i0}(\mathbf{z}_{i}(\nu,w),q_{0}(\nu),\nu,w)(i=1,2,...,N)$
		satisfies
		\[
		\frac{\partial\mathbf{z}_{i}(\nu,w)}{\partial\nu}A\nu=f_{i0}(\mathbf{z}_{i}(\nu,w),q_{0}(\nu),\nu,w)
		\]
		where $\mathbf{z}_{i}(\nu,w)$ is a smooth function with $\mathbf{z}_{i}(0,0)=0$.
	\end{fact}
	Meanwhile, under Assumption \ref{fact:5}, one can compute the solution
	to the regulator equation related to (\ref{eq:1}) and (\ref{eq:3})
	(see \cite{key-3,key-10}). The solution is given by:
	\[
	\mathbf{x}_{i1}(\nu)=q_{0}(\nu),
	\]
	\begin{align*}
	& \mathbf{x}_{i,j+1}(\nu,w)\\
	= & b_{ij}^{-1}(w)\left(\frac{\partial\mathbf{x}_{ij}}{\partial\nu}A\nu-f_{ij}(\mathbf{z}_{i},\mathbf{x}_{i1},...,\mathbf{x}_{ij},\nu,w)\right),
	\end{align*}
	\begin{align*}
	\mathbf{u}_{i}(\nu,w)=b_{in}^{-1}(w)\left(\frac{\partial\mathbf{x}_{in}}{\partial\nu}A\nu-f_{in}(\mathbf{z}_{i},\mathbf{x}_{i1},...,\mathbf{x}_{in},\nu,w)\right)
	\end{align*}
	where $j=1,2,...,n-1$. In addition, define $\mathbf{x}_{i,n+1}(\nu,w)\triangleq\mathbf{u}_{i}(\nu,w)$. 
	
	We also make the following standard assumption for $\mathbf{x}_{i1}(\nu),\mathbf{x}_{i,j+1}(\nu,w)(j=1,2,...,n)$.
	\begin{fact}
		\label{fact:Assume--are}\textcolor{black}{Assume that} $\mathbf{x}_{i1}(\nu),\mathbf{x}_{i,j+1}(\nu,w)(j=1,2,...,n)$
		are all polynomials in $\nu$ with coefficients depending on $w$.
	\end{fact}
	\begin{rem}
		If the considered system (\ref{eq:3}) is in a polynomial form, Assumption
		\ref{fact:Assume--are} will hold according to \cite{key-11aa}. Assumption
		3 guarantees the solvability of the output regulation problem. 
	\end{rem}
	By resorting to \cite{key-3,key-10}, Assumption \ref{fact:Assume--are}
	indicates that for any $\nu\in\mathbb{V},w\in\mathbb{W},j=1,2,...,n$,
	we have
	\[
	\frac{d^{\overline{n}_{ij}}\mathbf{x}_{i,j+1}}{dt^{\overline{n}_{ij}}}=\lambda_{i1}\mathbf{x}_{i,j+1}+\lambda_{i2}\frac{d\mathbf{x}_{i,j+1}}{dt}+\cdots+\lambda_{i\overline{n}_{ij}}\frac{d^{(\overline{n}_{ij}-1)}\mathbf{x}_{i,j+1}}{dt^{(\overline{n}_{ij}-1)}}
	\]
	where $\overline{n}_{ij}\in\mathbb{N}$. $\lambda_{i1},...,\lambda_{i\overline{n}_{ij}}$
	are real constants such that the roots of the polynomial $p_{ij}(s)=s^{\overline{n}_{ij}}-\lambda_{i1}-\lambda_{i2}s-\cdots-\lambda_{i\overline{n}_{ij}}s^{\overline{n}_{ij}-1}$
	are distinct with zero real parts.
	
	Then, given any column vector $N_{ij}\in\mathbb{R}^{\overline{n}_{ij}}$
	and Hurwitz matrix $M_{ij}$ satisfying $(M_{ij},N_{ij})$ are controllable,
	we have
	\[
	T_{ij}\Psi_{ij}-M_{ij}T_{ij}=N_{ij}\Gamma_{ij}
	\]
	where $T_{ij}$ is a nonsingular matrix. $\Gamma_{ij}=[1\thinspace0\thinspace...\thinspace0],$
	\[
	\Psi_{ij}=\left[\begin{array}{cccc}
	& 1\\
	&  & \ddots\\
	&  &  & 1\\
	\lambda_{i1} & \lambda_{i2} & \cdots & \lambda_{i\overline{n}_{ij}}
	\end{array}\right].
	\]
	
	Let $\theta_{ij}(\nu,w)=T_{ij}\mathrm{col}(\mathbf{x}_{i,j+1},\frac{d\mathbf{x}_{i,j+1}}{dt},...,\frac{d^{(\overline{n}_{ij}-1)}\mathbf{x}_{i,j+1}}{dt^{(\overline{n}_{ij}-1)}})$.
	Thus, we obtain: 
	\begin{align}
	\dot{\theta}_{ij}(\nu,w)= & T_{ij}\Psi_{ij}T_{ij}^{-1}\theta_{ij}(\nu,w),\nonumber \\
	{\color{black}\mathbf{x}_{i,j+1}(\nu,w)}= & {\color{black}\Phi_{ij}\theta_{ij}(\nu,w)}\label{eq:3-3}
	\end{align}
	where $\Phi_{ij}=\Gamma_{ij}T_{ij}^{-1}.$
	
	It can be seen that (\ref{eq:3-3}) generates the steady state $\mathbf{x}_{i,j+1}(\nu,w)$.
	Then we can design the following dynamic compensator:
	\begin{align}
	\dot{\eta}_{ij} & =M_{ij}\eta_{ij}+N_{ij}x_{i,j+1},\nonumber \\
	\dot{\eta}_{in} & =M_{in}\eta_{in}+N_{in}u_{i}\label{eq:4}
	\end{align}
	where $\eta_{ij},\eta_{in}$ are dynamic variables and $j=1,2,...,n-1$.
	
	(\ref{eq:4}) is also called the internal model for system (\ref{eq:1})
	and (\ref{eq:3}). It plays a pivotal role in solving the output regulation
	problem.
	
	Finally, the following change of coordinates is considered for system
	(\ref{eq:3}): 
	\begin{align}
	\overline{z}_{i} & =z_{i}-\mathbf{z}_{i}(\nu,w),\nonumber \\
	\overline{x}_{i1} & =x_{i1}-\mathbf{x}_{i1}(\nu),\nonumber \\
	{\color{black}\overline{x}_{ij}} & {\color{black}=x_{ij}-\Psi_{i,j-1}\eta_{i,j-1}\thinspace(j=2,...,n),}\nonumber \\
	{\color{black}{\color{black}\tilde{\eta}_{ij}}} & {\color{black}{\color{black}=\eta_{ij}-\theta_{ij}(\nu,w)-b_{ij}^{-1}(w)N_{ij}\overline{x}_{ij}}\thinspace(j=1,...,n),}\label{eq:6-1}\\
	\overline{x}_{i,n+1} & \triangleq\overline{u}_{i}=u_{i}-\Psi_{in}\eta_{in}.\label{eq:7}
	\end{align}
	
	Then, systems (\ref{eq:1}) and (\ref{eq:3}) can be written as:
	\begin{align}
	\dot{\overline{z}}_{i} & =\overline{f}_{i0}(\overline{z}_{i},\overline{x}_{1i},\nu,w),\nonumber \\
	\dot{\tilde{\eta}}_{ij} & =M_{ij}\tilde{\eta}_{ij}+g_{ij}(\overline{z}_{i},\tilde{\eta}_{i1},...,\tilde{\eta}_{i,j-1},\overline{x}_{i1},...,\overline{x}_{ij},\nu,w),\nonumber \\
	\dot{\overline{x}}_{ij} & =\overline{f}_{ij}(\overline{z}_{i},\tilde{\eta}_{i1},...,\tilde{\eta}_{ij},\overline{x}_{i1},...,\overline{x}_{ij},\nu,w)+b_{ij}(w)\overline{x}_{i,j+1},\nonumber \\
	e_{i} & =\overline{x}_{i1},\thinspace i=1,2,...,N;\thinspace j=1,2,...,n\label{eq:6}
	\end{align}
	where $\overline{f}_{i0}(\cdot),g_{ij}(\cdot),\overline{f}_{ij}(\cdot)(i=1,...,N;j=1,...,n)$
	are sufficiently smooth nonlinear functions with $\overline{f}_{i0}(0,...,0,\nu,w)=0,g_{ij}(0,...,0,\nu,w)=0,\overline{f}_{ij}(0,...,0,\nu,w)=0$
	for $\forall\nu\in\mathbb{V},\forall w\in\mathbb{W}$. 
	
	\textcolor{black}{It is noted that according to the above change of
		coordinates, the output regulation problem is transformed into the
		stabilization problem. Namely, if one can stabilize system (\ref{eq:6}),
		$i.e.$, find a controller $u_{i}$ to make $\overline{x}_{ij}(t)\rightarrow0(i=1,...,N;j=1,...,n)$
		as $t\rightarrow+\infty$, then the error $e_{i}$ will be regulated
		to zero. Therefore, in the following we will mainly consider the stabilization
		problem of system (\ref{eq:6}).}
	
	For the $\overline{z}_{i}$-system, we make the following assumption.
	\begin{fact}
		\label{fact:=00005B=00005D-Assume-there}\cite{key-11,key-21a} Assume
		that there exists a $C^{2}$ positive definite Lyapunov function $V_{i0}(\overline{z}_{i})$
		such that 
		\[
		\frac{\partial V_{i0}(\overline{z}_{i})}{\partial\overline{z}_{i}}\overline{f}_{i0}(\overline{z}_{i},0,\nu,w)\leq-\gamma_{i0}||\overline{z}_{i}||^{2}
		\]
		where $\gamma_{i0}$ is a known positive constant.
	\end{fact}
	\begin{rem}
		\textcolor{black}{The $z_{i}$-subsystem represents the dynamic uncertainty/unmodeled
			dynamics of the system. The states of the $z_{i}$-subsystem may not
			be available for feedback control. Assumption \ref{fact:=00005B=00005D-Assume-there}
			means that the zero dynamic of the $z_{i}$-system is asymptotically
			stable. It is less conservative than the assumption of input-to-state
			stability in \cite{key-11aa}. As a result, it is possible to find
			a control law that does not rely on the states of the $z_{i}$-subsystem.
			A lot of real practical systems satisfy Assumptions \ref{fact:Assume--are}
			and \ref{fact:=00005B=00005D-Assume-there}, such as Lorenz system,
			Chua's circuit, servo motors and robot mainpulators.}
	\end{rem}
	\textit{3) Useful results}
	
	We present some properties of matrix norms and useful inequalities.
	\begin{lem}
		\label{lem:1)-(Gronwall's-inequality)}
		
		1) (Property of skew-symmetric matrix) Given any skew-symmetric matrix
		$A\in\mathbb{R}^{n\times n}$ and matrix $B\in\mathbb{R}^{n\times m}$,
		we have $||\mathrm{e}^{A}||=1$ and $||\mathrm{e}^{A}B||=||B||$;
		
		2) \cite{key-22} For some square matrices $A,B\in\mathbb{R}^{n\times n}$,
		$||\mathrm{e}^{A}-\mathrm{e}^{B}||\leq\mathrm{e}^{||A||+||B-A||}||B-A||$;
		
		3) (Gronwall's inequality) Suppose
		\[
		u(t)\leq\rho_{1}+\int_{t_{0}}^{t}\rho_{2}u(\tau)d\tau
		\]
		for $\forall t\in[t_{0},+\infty)$ where $u(t):[t_{0},+\infty)\rightarrow\mathbb{R}$
		is a time-varying function, $\rho_{1},\rho_{2},t_{0}>0$ are positive
		constants. Then, $u(t)\leq\rho_{1}\mathrm{e}^{\rho_{2}(t-t_{0})}$;
		
		4) \cite{key-23} Given $u,u^{*}\in\mathbb{R}$ with $u\in[-R,R]$,
		then $|u-\mathrm{sat}_{R}(u^{*})|\leq\min\{|u-u^{*}|,2R\}$ where
		$R$ is a positive constant.
	\end{lem}
	
	\section{PET distributed observer}
	
\begin{figure}
		\begin{centering}
			\includegraphics[scale=0.70]{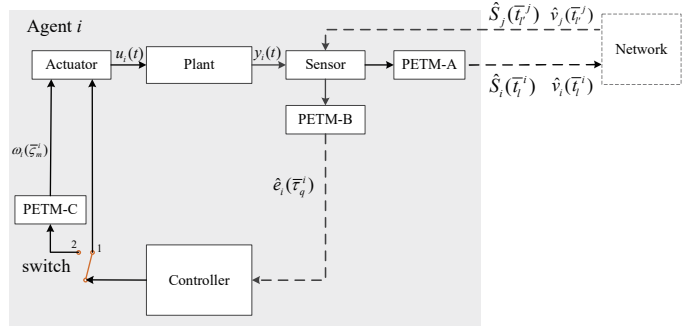}
			\par\end{centering}
		\caption{\label{fig:3-1-1}Event-triggered control scheme.}
	\end{figure}

	The proposed controller structure is illustrated in Fig. \ref{fig:3-1-1}
	(The switch is on node 1. Section IV-D will discuss the case when
	the switch is on node 2). It is composed of a PET distributed observer
	and a PET control law. The PET distributed observer is implemented
	in the sensor side to estimate the leader information. The control
	law uses estimated information to generate control signal. PET mechanisms
	are used for communications between each connected agent pair and
	the sensor-to-controller transmission channel in each agent. 
	
	Next, we will explain the PET distributed observer in this section,
	where two different cases are considered. The control law will be
	explained in the next section.

	\subsection{Case one}
	
	In this case, we assume that only a small number of the followers
	know the information $\nu$ of the leader. That is, only followers
	connected to the leader have access to $\nu$ (For instance, in Fig.
	\ref{fig:3-1-1-1}, among the four followers only agent 1 know $\nu$).
	Meanwhile, the matrix $A$ of the leader is known to all the followers.
	We design the following distributed observer for agent $i(i=1,2,...,N)$.
	\begin{align}
	\dot{\hat{\nu}}_{i} & =A\hat{\nu}_{i}+\mu_{2}\sum_{j=0}^{N}a_{ij}(\overline{\nu}_{j}(t,\overline{t}_{l'}^{j})-\overline{\nu}_{i}(t,\overline{t}_{l}^{i}))\label{eq:6-1-1-1}
	\end{align}
	\textcolor{black}{where $\hat{\nu}_{i}$ is used to estimate the leader
		information $\nu$,}
	\begin{equation}
	\overline{\nu}_{i}(t,\overline{t}_{l}^{i})=\mathrm{e}^{A(t-\overline{t}_{l}^{i})}\hat{\nu}_{i}(\overline{t}_{l}^{i})(i=1,...,N)\label{eq:7-2-1-1}
	\end{equation}
	with $\hat{\nu}_{0}\triangleq\nu$, $\overline{\nu}_{0}(t,t_{l}^{0})=\mathrm{e}^{A(t-t_{l}^{0})}\hat{\nu}_{0}(t_{l}^{0})=\mathrm{e}^{A(t-t_{l}^{0})}\nu(t_{l}^{0})=\nu(t)$.
	$\mu_{2}>0$ is a positive parameter.
	
	\textcolor{black}{The above distributed observer (\ref{eq:6-1-1-1})
		runs with respect to the time $t\in[0,+\infty)$. }Next, we will explain
	the time instants $\overline{t}_{l}^{i}$ and $\overline{t}_{l'}^{j}$.
	Let $0=t_{0}^{i}<t_{1}^{i}<\cdots<t_{k}^{i}<\cdots$ denote the sampling
	time instants for agent $i$ where $t_{k}^{i}\triangleq kT^{i}$ and
	$T^{i}>0$ represents the sampling period. Let $T\triangleq\underset{i\in\{1,2,...,N\}}{\max}T^{i}$
	and define set $\Omega_{T}^{i}=\{t_{0}^{i},t_{1}^{i},...,t_{k}^{i},...\}.$
	With slight abuse of notation, we use $t_{k}^{i}$ and $t_{k'}^{j}$
	denote the latest sampling time instants for agent $i$ and $j$ at
	the current time $t$.
	
	Then, let $0=\overline{t}_{0}^{i}<\overline{t}_{1}^{i}<\cdots<\overline{t}_{l}^{i}<\cdots$
	denote the event-triggered time instants. On time instant $\overline{t}_{l}^{i}$,
	agent $i$ will send $\hat{\nu}_{i}(\overline{t}_{l}^{i})$ to its
	neighbors. $\overline{t}_{l}^{i}$ is determined by the Periodic Event-Triggered
	Mechanism A (PETM-A) in Fig. \ref{fig:3-1-1}, that is
	\begin{equation}
	\overline{t}_{l+1}^{i}=\mathrm{inf}\{\tau>\overline{t}_{l}^{i}|\tau\in\Omega_{T}^{i},h_{\nu}^{i}(\tau,\overline{t}_{l}^{i})>0\}\label{eq:26-5}
	\end{equation}
	where 
	\[
	h_{\nu}^{i}(\tau,\overline{t}_{l}^{i})=||\hat{\nu}_{i}(\tau)-\overline{\nu}_{i}(\tau,\overline{t}_{l}^{i})||-\iota_{\nu}\mathrm{e}^{-\gamma_{\nu}\tau}
	\]
	with positive constants $\iota_{\nu},\gamma_{\nu}>0$.
	
	It can be seen that the set $\Omega_{ET}^{i}\triangleq\{\overline{t}_{0}^{i},\overline{t}_{1}^{i},...,\overline{t}_{l}^{i},...\}\subseteq\Omega_{T}^{i}.$
	Also let $\overline{t}_{l}^{i}$ and $\overline{t}_{l'}^{j}$ denote
	the latest event-triggered time instants for agent $i$ and $j$ on
	$[t_{k}^{i},t_{k+1}^{i})$ and $[t_{k'}^{j},t_{k'+1}^{j})$ respectively.
	Then, we have:
	\begin{thm}
		\label{thm:1-1}\textcolor{black}{Given the multi-agent systems with
			the leader (\ref{eq:1}) and the PET distributed observer (\ref{eq:6-1-1-1}),
			there exists a sufficiently small $T$ such that $\tilde{\nu}_{i}\triangleq\hat{\nu}_{i}-\nu(i=1,2,...,N)$
			converges to zero exponentially. Moreover, $||\tilde{\nu}_{i}||\in\mathsf{E}(\gamma_{\nu})$
			for $\forall i=1,2,...,N$.}
	\end{thm}
	\begin{IEEEproof}
		The proof is given in Appendix B. It is based on the properties of
		time-delay systems, exponential functions, and matrix norms.
	\end{IEEEproof}
	\begin{rem}
		\textcolor{black}{There are two time sequences for each agent $i$.
			That is the sampling time instants $t_{k}^{i}(k=0,1,2,...)$ and the
			event-triggered time instants $\overline{t}_{l}^{i}(l=0,1,2,...)$.
			Here, we want to emphasize that even though the sampling period $T^{i}=t_{k+1}^{i}-t_{k}^{i}>0$
			may be small, the inter-event time }\textit{\textcolor{black}{$\overline{t}_{l+1}^{i}-\overline{t}_{l}^{i}$}}\textcolor{black}{{}
			can be large. In fact, from Theorem 1, we know $\tilde{\nu}_{i}\triangleq\hat{\nu}_{i}-\nu$
			will converges to zero if $T^{i}=t_{k+1}^{i}-t_{k}^{i}$ is small
			enough. There is no special requirement for }\textit{\textcolor{black}{$\overline{t}_{l+1}^{i}-\overline{t}_{l}^{i}$}}\textcolor{black}{{}
			except for the event-triggered mechanism (\ref{eq:26-5}). The inter-event
			time }\textit{\textcolor{black}{$\overline{t}_{l+1}^{i}-\overline{t}_{l}^{i}$}}\textcolor{black}{{}
			can be made large by increasing the threshold in the event-triggered
			condition (\ref{eq:26-5}) (see the simulation in Section V in the
			supplmentary file).}
		
		\textcolor{black}{Based on the proofs in Appendix B and Lemma \ref{lem:Consider-the-following-1}
			in Appendix A, we know when $T$ satisfies the following inequality,
			\begin{align}
			T<\mathrm{min} & \left\{ \frac{1}{6\mu_{2}||\mathcal{H}||^{2}}-\frac{\gamma_{\nu}||P||}{3\mu_{2}^{2}||\mathcal{H}||^{2}}\right.,\nonumber \\
			& \left.\frac{1}{\mu_{2}||P\mathcal{H}||+3\mu_{2}^{2}||P\mathcal{H}||+\gamma_{\nu}}\right\} \label{eq:45-1}
			\end{align}
			where $P$ is a positive definite matrix such that $P\mathcal{H}+\mathcal{H}^{\mathrm{T}}P=2I$,
			we have $||\tilde{\nu}_{i}||\in\mathsf{E}(\gamma_{\nu})$ for $\forall i=1,2,...,N$.
			By (\ref{eq:45-1}), we can see that by decreasing the values of $\mu_{2},\gamma_{\nu}$.
			the sampling period $T$ can be increased. This also implies that the
			communication burden can be reduced. }
	\end{rem}
	
	\subsection{Case two}
	
	In this case, we assume that the state $\nu$ and matrix $A$ of the
	leader are known by a portion of the followers. Then, we design the
	following distributed observer for agent $i(i=1,2,...,N)$.
	\begin{align}
	\dot{\hat{A}}_{i} & =\mu_{1}\sum_{j=0}^{N}a_{ij}(\hat{A}_{j}(\overline{t}_{l'}^{j})-\hat{A}_{i}(\overline{t}_{l}^{i})),\label{eq:5-1-1}\\
	\dot{\hat{\nu}}_{i} & =\hat{A}_{i}(\overline{t}_{l}^{i})\hat{\nu}_{i}+\mu_{2}\sum_{j=0}^{N}a_{ij}(\overline{\nu}_{j}(t,\overline{t}_{l'}^{j})-\overline{\nu}_{i}(t,\overline{t}_{l}^{i}))\label{eq:6-1-1}
	\end{align}
	where \textcolor{black}{$\hat{\nu}_{i},\hat{A}_{i}$ are used to estimate
		the leader information $\nu,A$,}
	\begin{equation}
	{\color{black}\overline{\nu}_{i}(t,\overline{t}_{l}^{i})=\mathrm{e}^{\hat{A}_{i}(\overline{t}_{l}^{i})(t-\overline{t}_{l}^{i})}\hat{\nu}_{i}(\overline{t}_{l}^{i})(i=1,...,N)}\label{eq:7-2-1}
	\end{equation}
	with $\hat{\nu}_{0}\triangleq\nu$, $\overline{\nu}_{0}(t,t_{l}^{0})=\mathrm{e}^{A(t-t_{l}^{0})}\hat{\nu}_{0}(t_{l}^{0})=\mathrm{e}^{A(t-t_{l}^{0})}\nu(t_{l}^{0})=\nu(t)$.
	$\mu_{1},\mu_{2}$ are positive parameters. 
	
	$\overline{t}_{l}^{i}$ and $\overline{t}_{l'}^{j}$ are event-triggered
	time instants similar to the case in Section III-A. They are determined
	by the following PET mechanism:
	\begin{equation}
	\overline{t}_{l+1}^{i}=\mathrm{inf}\{\tau>\overline{t}_{l}^{i}|\tau\in\Omega_{T}^{i},h_{A}^{i}(\tau,\overline{t}_{l}^{i})>0,h_{\nu}^{i}(\tau,\overline{t}_{l}^{i})>0\}\label{eq:26}
	\end{equation}
	where 
	\[
	h_{A}^{i}(\tau,\overline{t}_{l}^{i})=||\hat{A}_{i}(\tau)-\hat{A}_{i}(\overline{t}_{l}^{i})||-\iota_{A}\mathrm{e}^{-\gamma_{A}\tau},
	\]
	\[
	h_{\nu}^{i}(\tau,\overline{t}_{l}^{i})=||\hat{\nu}_{i}(\tau)-\overline{\nu}_{i}(\tau,\overline{t}_{l}^{i})||-\iota_{\nu}\mathrm{e}^{-\gamma_{\nu}\tau}
	\]
	with positive constants $\iota_{A},\iota_{\nu},\gamma_{A},\gamma_{\nu}>0$.
	
	Now we present our second result.
	\begin{thm}
		\label{thm:1}\textcolor{black}{Given the multi-agent systems with
			the leader (\ref{eq:1}) and the PET distributed observer (\ref{eq:5-1-1})-(\ref{eq:6-1-1}),
			there exists a sufficiently small $T$ such that $\tilde{A}_{i}\triangleq\hat{A}_{i}-A$
			and $\tilde{\nu}_{i}\triangleq\hat{\nu}_{i}-\nu(i=1,2,...,N)$ converge
			to zero exponentially. Moreover, $||\tilde{A}_{i}||\in\mathsf{E}(\gamma_{A})$
			and $||\tilde{\nu}_{i}||\in\mathsf{E}(\mathrm{min}(\gamma_{A},\gamma_{\nu}))$
			for $\forall i=1,2,...,N$.}
	\end{thm}
	\begin{IEEEproof}
		The proof is also put in Appendix B.
	\end{IEEEproof}
	\begin{rem}
		For the proposed distributed PET observer, the data transmission and
		PET condition are required to be monitored only periodically. Thus,
		the Zeno behavior is excluded naturally because a minimum positive
		constant exists between triggered time instants, $i.e.$, $\overline{t}_{l+1}^{i}-\overline{t}_{l}^{i}\geq T^{i}$.
		Meanwhile, compared with our previous work \cite{key-11a}, the proposed
		method has several essential differences: 1) The communication among
		various agents is asynchronous because each agent $i$ has a different
		sampling time $T^{i}$. This makes the proof of Theorems \ref{thm:1-1}
		and \ref{thm:1} quite different from \cite{key-11a}; 2) The convergence
		rate for the observer is provided, which will be used in the stability
		analysis of the PET controller in Section IV. 
	\end{rem}
	\begin{rem}
		Evidently, the distributed observer in Section III-B is more general
		than that in Section III-A. It can be used for more complex environment.
		In the following controller design in Section IV, we assume that the
		matrix $A$ is known, $i.e.,$ the observer in Section III-A is used.
		The application of the distributed observer in Section III-B is similar
		but out of the scope of this study. One can resort to \cite{key-8,key-9}
		for more information. This observer can be used in linear multi-agent
		systems, multiple Euler-Lagrange systems $etc$.
	\end{rem}
	
	\section{PET output feedback controller}
	
	We will consider the design of PET output feedback controller for
	the nonlinear multi-agent systems given by (\ref{eq:1})-(\ref{eq:3})
	in this section. The design will be divided into the following steps.
	
	\subsection{System transformation}
	
	\textcolor{black}{From Section II-B, it can be seen that the considered
		Problem 1 can be solved if system (\ref{eq:6}) is stabilized. That is
		we can design a controller $u_{i}(i=1,2,...,N)$ for (\ref{eq:6})
		such that all the states $\overline{x}_{ij}\rightarrow0(i=1,2,...,N;j=1,2,...,n)$.
		However, it is not easy to find such a controller $u_{i}$ since only
		the output $e_{i}=\overline{x}_{i1}$ is measurable and  system
		(\ref{eq:6}) is in a strict feedback form. In this subsection, we
		will introduce a coordinate transformation technique for system (\ref{eq:6}).
		This transformation is useful for the subsequent output feedback controller
		design.}
	
	\textcolor{black}{Using (\ref{eq:6}), define}
	\begin{align}
	\xi_{i1} & \triangleq\overline{x}_{i1},\nonumber \\
	\xi_{i2} & \triangleq\dot{\xi}_{i1}=\overline{f}_{i1}(\overline{z}_{i},\tilde{\eta}_{i1},\overline{x}_{i1},\nu,w)+b_{i1}(w)\overline{x}_{i2},\label{eq:22-3}\\
	\xi_{ij} & \triangleq\dot{\xi}_{i,j-1}(j=3,4,...,n+1).\nonumber 
	\end{align}
	
	Note that $\xi_{ij}$ has the following properties:
	\begin{prop}
		\label{prop:1}For $i=1,2,...,N$; $j=1,2,...,n$,
		\begin{align}
		\xi_{ij} & =\overline{\xi}_{ij}(\overline{z}_{i},\tilde{\eta}_{i1},...,\tilde{\eta}_{i,j-1},\overline{x}_{i1},...,\overline{x}_{ij},\nu,w),\label{eq:11-2}\\
		\overline{x}_{ij} & =\chi_{ij}(\overline{z}_{i},\tilde{\eta}_{i1},...,\tilde{\eta}_{i,j-1},\xi_{i1},...,\xi_{ij},\nu,w).\label{eq:12-1}
		\end{align}
		Specifically, 
		\[
		\xi_{i,n+1}=\phi_{i}(\overline{z}_{i},\tilde{\eta}_{i1},...,\tilde{\eta}_{in},\xi_{i1},...,\xi_{in},\nu,w)+b_{in}(w)\overline{u}_{i}
		\]
		where $\overline{\xi}_{ij}(\cdot),\chi_{ij}(\cdot),\phi_{i}(\cdot)$
		are smooth functions with $\overline{\xi}_{ij}(0,...,0,\nu,w)=\chi_{ij}(0,...,0,\nu,w)=\phi_{i}(0,...,0,\nu,w)=0$. 
	\end{prop}
	\begin{IEEEproof}
		The proof is put in Appendix B.
	\end{IEEEproof}
	On the basis of this transformation, system (\ref{eq:6}) can be rewritten
	as follows: 
	\begin{align}
	\dot{\overline{z}}_{i} & =f_{i0}(\overline{z}_{i},\xi_{i1},\nu,w),\nonumber \\
	\dot{\tilde{\eta}}_{ij} & =M_{ij}\tilde{\eta}_{ij}+h_{ij}(\cdot),\thinspace j=1,2,...,n\nonumber \\
	\dot{\xi}_{i1} & =\xi_{i2},\nonumber \\
	\dot{\xi}_{ij} & =\xi_{i,j+1},\thinspace j=2,...,n-1\label{eq:22-2}\\
	\dot{\xi}_{in} & =\phi_{i}(\overline{z}_{i},\tilde{\eta}_{i1},...,\tilde{\eta}_{in},\xi_{i1},...,\xi_{in},\nu,w)+b_{in}(w)\overline{u}_{i},\nonumber \\
	e_{i} & =\xi_{i1},\thinspace i=1,2,...,N\nonumber 
	\end{align}
	where $h_{ij}(\cdot)=h_{ij}(\overline{z}_{i},\tilde{\eta}_{i1},...,\tilde{\eta}_{i,j-1},\xi_{i1},...,\xi_{ij},\nu,w)$
	is a smooth function with $h_{ij}(0,...,0,\nu,w)=0$. 
	
	Next, inspired by the backstepping technique, let
	\begin{align}
	\zeta_{i1} & =\xi_{i1},\nonumber \\
	\zeta_{ij} & =\xi_{ij}-\alpha_{i,j-1}(j=2,3,...,n)\label{eq:24-1}
	\end{align}
	where
	\begin{equation}
	\alpha_{ij}=-Q_{ij}\zeta_{ij},\thinspace j=1,2,...,n-1\label{eq:22-4}
	\end{equation}
	with a positive design parameter $Q_{ij}>0$.
	
	Using (\ref{eq:24-1}), (\ref{eq:22-2}) becomes 
	\begin{align}
	\dot{\overline{z}}_{i} & =f_{i0}(\overline{z}_{i},\zeta_{i1},\nu,w),\nonumber \\
	\dot{\tilde{\eta}}_{ij} & =M_{ij}\tilde{\eta}_{ij}+\overline{h}_{ij}(\cdot),\thinspace j=1,...,n\nonumber \\
	\dot{\zeta}_{i1} & =\zeta_{i2}+\alpha_{i1},\nonumber \\
	\dot{\zeta}_{ij} & =\zeta_{i,j+1}+\alpha_{ij}-\dot{\alpha}_{i,j-1},\thinspace j=2,...,n-1\label{eq:22-1}\\
	\dot{\zeta}_{in} & =\overline{\phi}_{i}(\overline{z}_{i},\tilde{\eta}_{i1},...,\tilde{\eta}_{in},\zeta_{i1},...,\zeta_{in},\nu,w)-\dot{\alpha}_{i,n-1}+b_{in}(w)\overline{u}_{i},\nonumber \\
	e_{i} & =\zeta_{i1},\thinspace i=1,2,...,N\nonumber 
	\end{align}
	where $\overline{h}_{ij}(\cdot)=\overline{h}_{ij}(\overline{z}_{i},\tilde{\eta}_{i1},...,\tilde{\eta}_{i,j-1},\zeta_{i1},...,\zeta_{ij},\nu,w),\overline{\phi}_{i}(\cdot)$
	are smooth functions with $\overline{h}_{ij}(0,...,0,\nu,w)=\overline{\phi}_{i}(0,...,0,\nu,w)=0$.
	Meanwhile, $\dot{\alpha}_{ij}$ has the following property:
	\begin{prop}
		\label{prop:2}\textcolor{black}{For $i=1,2,...,N$; $j=1,2,...,n-1$,}
		there exists a positive constant $\vartheta_{ij}(Q_{i1},Q_{i2},...,Q_{ij})$
		related with $Q_{i1},Q_{i2},...,Q_{ij}$ such that $|\dot{\alpha}_{ij}|\leq\vartheta_{ij}(Q_{i1},Q_{i2},...,Q_{ij})(|\zeta_{i1}|+\cdots+|\zeta_{ij}|+|\zeta_{i,j+1}|)$.
	\end{prop}
	\begin{IEEEproof}
		See Appendix C for detailed information.
	\end{IEEEproof}
	\textcolor{black}{It is noted that if one can find a controller $\overline{u}_{i}$
		to stabilize the transformed system (\ref{eq:22-2}) or (\ref{eq:22-1}),
		then (\ref{eq:6}) can be also stabilized. That is if $\xi_{ij}\rightarrow0$
		or $\zeta_{ij}\rightarrow0$, then $\overline{x}_{ij}\rightarrow0(i=1,...,N;j=1,...,n)$.
		Hence, in the following we will consider the stabilization problem
		of (\ref{eq:22-2}) and (\ref{eq:22-1}).}
	
	\subsection{State feedback controller}
	
	We will introduce a state feedback controller for system (\ref{eq:22-1})
	laying the foundation for the design of output feedback controller
	in the next subsection. The following Lyapunov function is considered:
	\begin{align}
	V_{i} & =\frac{V_{i0}(\overline{z}_{i})}{L_{i0}}+\sum_{j=1}^{n}\frac{\tilde{\eta}_{ij}^{\mathrm{T}}P_{ij}\tilde{\eta}_{ij}}{L_{ij}}+\sum_{j=1}^{n}\frac{1}{2}\zeta_{ij}^{2}\label{eq:24}
	\end{align}
	where $i=1,...,N$, $V_{i0}(\overline{z}_{i})$ is given in Assumption
	\ref{fact:=00005B=00005D-Assume-there}, $P_{ij}>0$ are positive
	definite matrices such that
	\[
	P_{ij}M_{ij}+M_{ij}^{\mathrm{T}}P_{ij}\leq-\beta_{ij}I
	\]
	where $\beta_{ij}>0$ is a positive constant. Because $M_{ij}$ is
	Hurwitz, $P_{ij}$ exists. $L_{i0},L_{ij}\geq1$ are scaling gains
	which will be explained in the proof of Lemma \ref{lem:For-system-(),}.
	
	Let $X_{i}=\mathrm{col}(\overline{z}_{i},\tilde{\eta}_{i1},...,\tilde{\eta}_{in},\zeta_{i1},...,\zeta_{in})$.
	Assume
	\[
	X_{i}(0)\in B_{r}\triangleq[r,-r]^{n_{X_{i}}}
	\]
	where $r$ is a positive constant, $n_{X_{i}}$ denotes the dimension
	of $X_{i}$.
	
	Then, there exists a constant $\overline{R}>0$ such that 
	\[
	V_{i}(X_{i})\leq\overline{R}
	\]
	for $\forall X_{i}\in B_{r}.$
	
	Next, define the following set
	\begin{align}
	\Omega_{R} & =\{X_{i}|V_{i}(X_{i})\leq R\triangleq\overline{R}+\Delta_{R}\}\label{eq:25}
	\end{align}
	where $\Delta_{R}>0$ is a positive design parameter which will be
	explained later.
	
	Then, we have:
	\begin{lem}
		\label{lem:For-system-(),}For system (\ref{eq:22-1}), \textcolor{black}{suppose
			$X_{i}(0)\in B_{r}$ and belongs to the set $\Omega_{R}$,} then there
		exists a virtual state feedback control effort $\overline{u}_{i}^{*}$
		given by
		\begin{equation}
		\overline{u}_{i}^{*}=K_{i}(R)\zeta_{in}\label{eq:26-2}
		\end{equation}
		such that
		\begin{align}
		\dot{V}_{i}\leq & -\frac{\tilde{\gamma}_{i}}{2}||\overline{z}_{i}||^{2}-\frac{\tilde{\varrho}_{i}}{2}\sum_{k=1}^{n}||\tilde{\eta}_{ij}||^{2}-\frac{1}{4}\sum_{j=1}^{n}\zeta_{ij}^{2}\nonumber \\
		& +\zeta_{in}b_{in}(w)(\overline{u}_{i}-\overline{u}_{i}^{*})\label{eq:27-3}
		\end{align}
		\textcolor{black}{where $K_{i}(R)$ is a sufficiently large control
			gain related with $R$, and $\tilde{\gamma}_{i},\tilde{\varrho}_{i}$
			are positive constants.}
	\end{lem}
	\begin{IEEEproof}
		See Appendix D.
	\end{IEEEproof}
	
	\subsection{Output feedback controller}
	
	First, a new PET high gain observer is proposed to estimate the transformed
	variable $\xi_{ij}$ and $\zeta_{ij}$ in (\ref{eq:22-3}) and (\ref{eq:24-1}).
	\textcolor{black}{Denote the sampling time instants as $0=\tau_{0}^{i}<\tau_{1}^{i}<\cdots<\tau_{p}^{i}<\cdots$.
		Let $\mathcal{T}^{i}=\tau_{p+1}^{i}-\tau_{p}^{i}$ denote the sampling
		period. Note that the sampling time instants can be asynchronous with
		the distributed observer developed in Section III. Also define set
		${\color{black}\Omega_{\mathcal{T}}^{i}=\{\tau_{0}^{i},\tau_{1}^{i},...,\tau_{p}^{i},...\}.}$}
	Meanwhile, the PET instants are denoted as: $0=\overline{\tau}_{0}^{i}<\overline{\tau}_{1}^{i}<\cdots<\overline{\tau}_{q}^{i}<\cdots$
	. Then the PET high gain observer is given by:
	\begin{align}
	\dot{\hat{\xi}}_{i1}= & \hat{\xi}_{i2}+\Gamma_{i}d_{1}(\hat{e}_{i}(\overline{\tau}_{q}^{i})-\hat{\xi}_{i1}),\nonumber \\
	\dot{\hat{\xi}}_{i2}= & \hat{\xi}_{i3}+\Gamma_{i}^{2}d_{2}(\hat{e}_{i}(\overline{\tau}_{q}^{i})-\hat{\xi}_{i1}),\nonumber \\
	\vdots\label{eq:27}\\
	\dot{\hat{\xi}}_{in}= & \hat{b}_{in}\overline{u}_{i}+\Gamma_{i}^{n}d_{n}(\hat{e}_{i}(\overline{\tau}_{q}^{i})-\hat{\xi}_{i1})\nonumber 
	\end{align}
	where $\hat{e}_{i}(t)=x_{i1}(t)-q_{0}(\hat{\nu})$, $\Gamma_{i}\geq1$,
	$\hat{b}_{in},d_{j}(j=1,2,...,n)>0$ are positive design parameters.\textcolor{black}{{}
		$d_{j}$ are the coefficients of some Hurwitz polynomial $p_{d}(s)=s^{n}+d_{1}s^{n-1}+\cdots+d_{n-1}s+d_{n}$.}
	
	The PET time instants are determined by the Periodic Event-Triggered
	Mechanism B (PETM-B) in Fig. \ref{fig:3-1-1}, that is
	\begin{equation}
	\overline{\tau}_{q+1}^{i}=\mathrm{inf}\{s>\overline{\tau}_{q}^{i}|s\in\Omega_{\mathcal{T}}^{i},h_{e}^{i}(s,\overline{\tau}_{q}^{i})>0\}\label{eq:26-1}
	\end{equation}
	where $h_{e}^{i}(s,\overline{\tau}_{q}^{i})=|\hat{e}_{i}(s)-\hat{e}_{i}(\overline{\tau}_{q}^{i})|-\iota_{e}|\hat{e}_{i}(s)|$
	with a positive constant $\iota_{e}$.
	
	Then the estimated values $\hat{\zeta}_{i1},\hat{\zeta}_{ij}(j=2,...,n)$
	are computed by\textcolor{black}{
		\begin{align*}
		\hat{\zeta}_{i1} & =\hat{\xi}_{i1},\\
		\hat{\zeta}_{ij} & =\hat{\xi}_{ij}-\hat{\alpha}_{i,j-1}(j=2,3,...,n),
		\end{align*}
		where
		\[
		\hat{\alpha}_{ij}=-Q_{ij}\hat{\zeta}_{ij},\thinspace j=1,2,...,n.
		\]
	}Based on the above estimation, $\overline{u}_{i}$ in (\ref{eq:22-1})
	is given by
	\begin{equation}
	\overline{u}_{i}=\mathrm{sat}_{\mathcal{R}}(K_{i}(R)\hat{\zeta}_{in})\label{eq:29-2-1}
	\end{equation}
	where $\mathcal{R}$ is a positive design parameter. $K_{i}(R)>0$
	is a control gain related with $R$.
	
	By (\ref{eq:7}), the actual control effort is computed as:
	\begin{align}
	u_{i}(t)= & \mathrm{sat}_{\mathcal{R}}(K_{i}(R)\hat{\zeta}_{in})+\Psi_{in}\eta_{in},\label{eq:26-4}\\
	\dot{\eta}_{in}= & M_{in}\eta_{in}+N_{in}u_{i}.\label{eq:27-2}
	\end{align}
	
	Our third result is as follows.
	\begin{thm}
		\label{thm:1-1-1}\textcolor{black}{Consider the multi-agent systems
			(\ref{eq:1})-(\ref{eq:3}) with the output feedback control controller
			(\ref{eq:26-4})-(\ref{eq:27-2}), PET high gain observer (\ref{eq:27})
			and PET distributed observer (\ref{eq:5-1-1})-(\ref{eq:6-1-1}).
			Suppose the initial states $X_{i}(0)\in B_{r}$ and belong to the
			set $\Omega_{R}$. Then, there exist a sufficiently large control
			gain $K_{i}(R)$ and sufficiently small sampling time periods $\mathcal{T}^{i},T^{i}$
			such that Problem \ref{prob:Given-a-multi-agent} is solvable.}
	\end{thm}
	\begin{IEEEproof}
		The proof is put in Appendix E.
	\end{IEEEproof}
	\begin{rem}
		The main result and its proof show that there exist sufficiently large
		control gains and small sampling periods such that the cooperative
		semi-global output regulation problem can be solved. Moreover, the
		controller (\ref{eq:26-4})-(\ref{eq:27-2}) is not complex and easy
		to be implemented. The detailed tuning method for the control gains
		and sampling times is out of the scope of this study. This is a common
		case for semi-global control problems as shown in \cite{key-11,key-23,key-24,key-25}.
		In addition, since the considered system (\ref{eq:3}) may contain
		some unknown nonlinearities such as $f_{i0}(z_{i},x_{i1},\nu,w)$,
		it is not easy to explicitly give the upper bound for sampling periods
		like \cite{key-11a,key-27}. 
		
		Some guidelines for the selections of the control parameters are as
		follows: Larger control gains can result in rapid response but serious
		oscillations. Smaller sampling period is beneficial for the stability
		of the system but may result in more communication burden. Increasing
		the parameters $\iota_{e},\iota_{\nu}$ and decreasing $\gamma_{\nu}$
		in the event triggered condition (\ref{eq:26-1}) and (\ref{eq:26-5})
		can result in a light communication burden but deteriorate the control
		performance.
		
		It is also noted that from the simulation results in Section V, we
		can see that the tuning of the control parameters is not tedious.
		One can first select a small sampling period and then gradually increase
		the control gains. It is not hard to stabilize the closed loop systems.
		Moreover, the simulation shows that the controller has strong robustness
		to the variations of sampling periods.
	\end{rem}
	
	\subsection{Extension}
	
	We give an extension to the proposed results. An extra PET mechanism
	is used between the controller and actuator in Fig. \ref{fig:3-1-1}.
	That is the switch is on node 2. In this case, the actual control
	effort is given by
	\begin{align}
	u_{i}(t)= & \omega_{i}(\overline{\varsigma}_{m}^{i}),\thinspace t\in[\overline{\varsigma}_{m}^{i},\overline{\varsigma}_{m+1}^{i}),\label{eq:26-4-2}\\
	\omega_{i}(t)= & \mathrm{sat}_{\mathcal{R}}(K_{i}(R)\hat{\zeta}_{in})+\Psi_{in}\eta_{in},\label{eq:32}\\
	\dot{\eta}_{in}= & M_{in}\eta_{in}+N_{in}u_{i}\label{eq:27-2-2}
	\end{align}
	where $0=\overline{\varsigma}_{0}^{i}<\overline{\varsigma}_{1}^{i}<\cdots<\overline{\varsigma}_{m}^{i}<\cdots$
	are the PET time instants. On time instant $\overline{\varsigma}_{m}^{i}$,
	the controller will transmit $\omega_{i}(\overline{\varsigma}_{m}^{i})$
	to the actuator. They are given by:
	\begin{equation}
	\overline{\varsigma}_{m+1}^{i}=\mathrm{inf}\{\tau>\overline{\varsigma}_{m}^{i}|\tau\in\Omega_{\mathcal{T}}^{i},h_{\omega}^{i}(\tau,\overline{\varsigma}_{m}^{i})>0\}\label{eq:26-1-1-1}
	\end{equation}
	where $h_{\omega}^{i}(\tau,\overline{\varsigma}_{m}^{i})=|\omega_{i}(\tau)-\omega_{i}(\overline{\varsigma}_{m}^{i})|-\iota_{\omega}|\omega_{i}(\tau)|$
	with a constant $\iota_{\omega}\geq0$. 
	
	Then, we have the last result in this paper.
	\begin{thm}
		\label{thm:1-1-1-1}\textcolor{black}{Consider the multi-agent systems
			(\ref{eq:1})-(\ref{eq:3}) with the PET output feedback control controller
			(\ref{eq:26-4-2})-(\ref{eq:27-2-2}), PET high gain observer (\ref{eq:27})
			and PET distributed observer (\ref{eq:5-1-1})-(\ref{eq:6-1-1}).
			Suppose the initial states $X_{i}(0)\in B_{r}$ and belong to the
			set $\Omega_{R}$. Then, there exist a sufficiently large control
			gain $K_{i}(R)$ and sufficiently small sampling time periods $\mathcal{T}^{i},T^{i}$
			such that }
		
		\textcolor{black}{1) All the signals are semi-globally uniformly bounded
			for $\forall t\in[0,+\infty)$; and,}
		
		\textcolor{black}{2) The output regulation error $e_{i}(t)\triangleq y_{i}(t)-y_{0}(t)$
			satisfies $\underset{t\rightarrow+\infty}{\lim}|e_{i}(t)|\leq\delta_{i}(\iota_{e},\iota_{\omega},\mathcal{T}^{i})$
			$\forall i\in\{1,2,...,N\}$ where $\delta_{i}(\iota_{e},\iota_{\omega},\mathcal{T}^{i})$
			is an increasing function with $\delta_{i}(0,0,0)=0$.}
	\end{thm}
	\begin{IEEEproof}
		The proof is put in Appendix E.
	\end{IEEEproof}
	\begin{rem}
		From the proof in Appendix E, the detail expression of $\delta_{i}(\iota_{e},\iota_{\omega},\mathcal{T}^{i})$
		could be complex and may be conservative. This is a common case when
		adopting Lyapunov function method \cite{key-28,key-29}. However,
		according to the property of $\delta_{i}(\iota_{e},\iota_{\omega},\mathcal{T}^{i})$,
		we know the regulation error can be made arbitrary small by tuning
		the design parameters $\iota_{e},\iota_{\omega},\mathcal{T}^{i}$.
	\end{rem}
	
	\section{Simulations}
	
	A group of four Lorenz systems is considered as follows:
	\begin{align*}
	\dot{z}_{i1}= & g_{i1}z_{i1}+g_{i2}x_{i1},\\
	\dot{z}_{i2}= & g_{i3}z_{i2}+z_{i1}x_{i1},\\
	\dot{x}_{i1}= & g_{i4}z_{i1}+g_{i5}x_{i1}-z_{i1}z_{i2}+x_{i2},\\
	\dot{x}_{i2}= & g_{i6}z_{i1}+g_{i7}z_{i2}x_{i1}+u_{i},\\
	y_{i}= & x_{i1},\thinspace i=1,2,3,4
	\end{align*}
	where $g_{i1}=-10$, $g_{i2}=10$, $g_{i3}=-8/3$, $g_{i4}=1$, $g_{i5}=0$,
	$g_{i6}=0.2$. 
	
\begin{figure}
		\begin{centering}
			\includegraphics[scale=0.7]{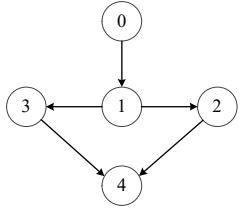}
			\par\end{centering}
		\caption{\label{fig:3-1-1-1}Communication graph.}
	\end{figure}

	The leader is given by
	\begin{align*}
	\dot{\nu} & =A\nu,\\
	y_{0} & =[1\thinspace0]\nu
	\end{align*}
	where $S=\left[\begin{array}{cc}
	0 & 1\\
	-1 & 0
	\end{array}\right].$ The communication graph is depicted in Fig. \ref{fig:3-1-1-1}.
	
	The control structure is composed of three parts, namely, the PET
	distributed observer (\ref{eq:5-1-1})-(\ref{eq:6-1-1}), the PET
	local observer (\ref{eq:27}) and the controller (\ref{eq:29-2-1}).
	The sampling time is set as $T^{1}=\mathcal{T}^{1}=0.01s$, $T^{2}=\mathcal{T}^{2}=0.015s$,
	$T^{3}=\mathcal{T}^{3}=0.02s$, $T^{4}=\mathcal{T}^{4}=0.025s$. The
	controller parameters of these three parts are set as $\mu=2$, $d_{1}=5$,
	$d_{2}=10$, $\varLambda_{i}=40$, $Q_{i1}=2$, $K_{i}=30(i=1,2,3,4)$.
	According to \cite{key-10}, $M_{i2},N_{i2}(i=1,2,3,4)$ in the controller
	(\ref{eq:27-2}) can be calculated as 
	\[
	M_{i2}=\left[\begin{array}{cccc}
	& 1\\
	&  & 1\\
	&  &  & 1\\
	-10 & -18 & -15 & -6
	\end{array}\right],N_{i2}=\left[\begin{array}{c}
	\\
	\\
	\\
	1
	\end{array}\right].
	\]
	
\begin{figure}
		\begin{centering}
			\includegraphics[scale=0.5]{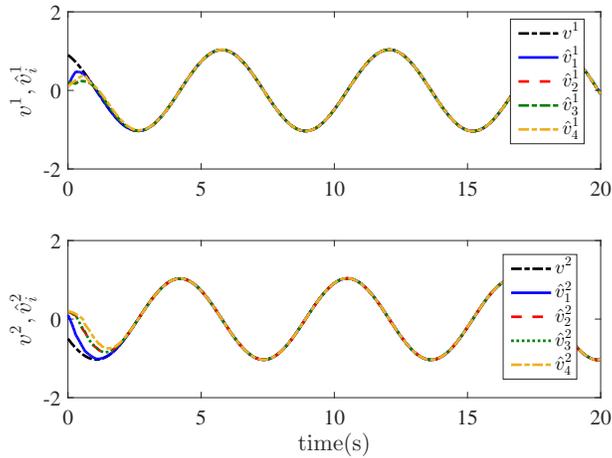}
			\par\end{centering}
		\caption{\label{fig:3-1-1-1-1}Performance of the distributed observer.}
	\end{figure}

\begin{figure}
		\begin{centering}
			\includegraphics[scale=0.5]{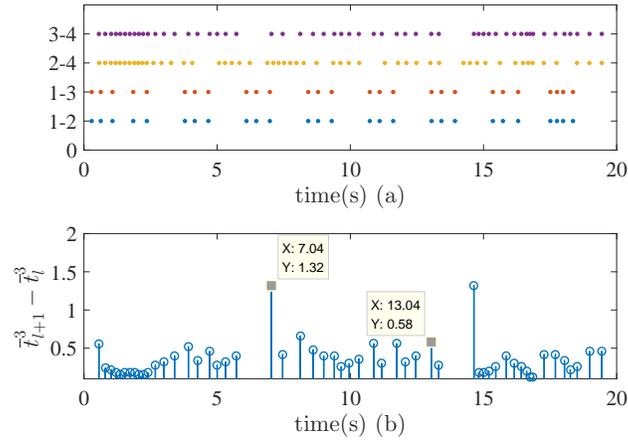}
			\par\end{centering}
		\caption{\label{fig:3-1-1-1-1-1}Event-triggered time instants.}
	\end{figure}

	The performance of the PET distributed observer is shown in Fig. \ref{fig:3-1-1-1-1}.
	The results demonstrate that each agent can estimate the information
	of the leader accurately. Fig. \ref{fig:3-1-1-1-1-1}(a) shows the
	event-triggered time instants between each agent pair. It can be seen
	that the communication burden has been reduced a lot. In addition,
	the communication of the multi-agent systems is asynchronous since
	the event-triggered time instants among different agents are different.
	\textcolor{black}{Fig. \ref{fig:3-1-1-1-1-1}(b) shows the inter-event
		times for agent 3. The inter-event times are much larger than the
		sampling period.} Meanwhile, they are multiples of the sampling time
	$\mathcal{T}^{i}$. This implies that not only the Zeno behavior is
	excluded, but also the data transmission is periodically triggered.
	\textcolor{black}{All these verify} the advantages of the developed
	distributed observer.
	
\begin{figure}
		\begin{centering}
			\includegraphics[scale=0.5]{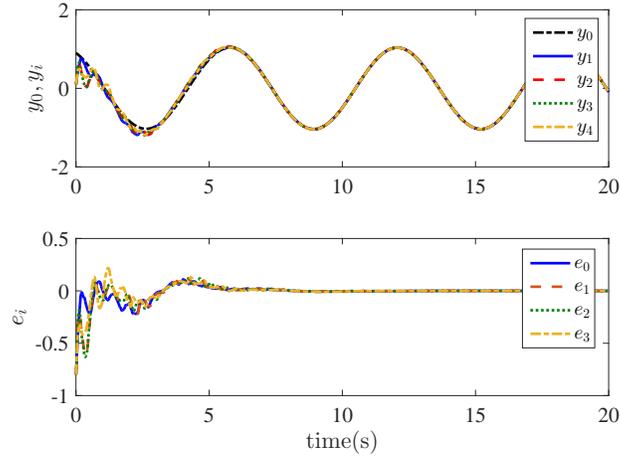}
			\par\end{centering}
		\caption{\label{fig:3-1-1-1-1-1-1}Control performance.}
	\end{figure}

	The control performance of the entire multi-agent systems is shown
	in Fig. \ref{fig:3-1-1-1-1-1-1}. It can be seen that the regulation
	error rapidly becomes zero in a very short time. Table I shows the
	event-triggered times. The table shows that the data transmission
	of the PET controller is much less than that of the sampled-data control
	strategy.
	\begin{table}
		\begin{centering}
			\caption{\label{tab:5}Event-triggered times for PETM-B under different $\iota_{e}$.}
			\par\end{centering}
		\centering{}%
		\begin{tabular}{cccc}
			\hline 
			& $\iota_{e}=0.05$ & $\iota_{e}=0.1$ & $\iota_{e}=0.2$\tabularnewline
			\hline 
			sampled data & 800 & 800 & 800\tabularnewline
			PET  & 390 & 324 & 271\tabularnewline
			\hline 
		\end{tabular}
	\end{table}

	\section{Conclusions}
	
	In this paper, the PET cooperative output regulation problem is considered
	for strict feedback nonlinear multi-agent systems. We propose a new
	PET distributed observer and a PET output feedback control law for
	this problem. The communication between various agents can be asynchronous.
	\textcolor{black}{Future works include considering PET output regulation
		for non-strict feedback nonlinear systems. For non-strict feedback
		nonlinear systems, the control effort and the states may be coupled
		with each other. This will make the problem more challenging.}
	
	\section{Appendix}
	
	\appendices{}
	
	\subsection{Two lemmas}
	
	In this section, we will present two key lemmas which will be used
	the in the proof of Theorems 1-2.
	\begin{lem}
		\label{lem:Consider-the-following}Consider the following system
		\begin{align}
		\dot{x}= & -\mu\Lambda_{1}\boldsymbol{x}_{d}+\mu\Lambda_{2}\boldsymbol{x}_{d}+\Lambda_{3}x+\mu\Lambda_{4}+\Lambda_{5}\label{eq:26-3}
		\end{align}
		where $x=(x_{1},x_{2},...,x_{N})^{\mathrm{T}}\in\mathbb{R}^{N}$,
		$\boldsymbol{x}_{d}=(x_{1}(t-d_{1}(t)),x_{2}(t-d_{2}(t)),...,x_{N}(t-d_{N}(t)))^{\mathrm{T}}\in\mathbb{R}^{N}$.
		$d_{i}(t)(i=1,2,...,N)\in\mathbb{R}$ are time-varying delays such
		that $0<d_{i}(t)\leq T$ with a positive constant $T$. $\mu>0$ is
		a positive constant. If $-\Lambda_{1}$ is a Hurwitz matrix and $\Lambda_{j}(j=2,3,4,5)\in\mathsf{E}(\gamma)$
		with a positive constant $\gamma$, then there exist a sufficiently
		large \textup{$\mu$} and small \textup{$T$} such that $x\in\mathsf{E}(\gamma)$.
	\end{lem}
	\begin{IEEEproof}
		(\ref{eq:26-3}) can be written as:
		\begin{align}
		\dot{x}= & -\mu\Lambda_{1}x+\mu\Lambda_{2}x+\mu\Lambda_{1}\eta-\mu\Lambda_{2}\eta\nonumber \\
		& +\Lambda_{3}x+\mu\Lambda_{4}+\Lambda_{5}\label{eq:27-4}
		\end{align}
		where $\eta(t)=\mathrm{col}(\eta_{1}(t),\eta_{2}(t),...,\eta_{N}(t))$
		with $\eta_{i}(t)=\int_{t-d_{i}(t)}^{t}\dot{x}_{i}(s)ds$.
		
		Consider the following Lyapunov-Krasovskii function
		\begin{equation}
		V=\frac{1}{2}x^{\mathrm{T}}Px+\int_{t-T}^{t}(s-t+T)||\dot{x}(s)||^{2}ds\label{eq:13-1}
		\end{equation}
		where $P$ is a positive definite matrix such that $P\Lambda_{1}+\Lambda_{1}^{\mathrm{T}}P=2I$.
		
		Using (\ref{eq:27-4}), the derivative of $V$ is computed as:
		\begin{align*}
		\dot{V}\leq & x^{\mathrm{T}}P(-\mu\Lambda_{1}x+\mu\Lambda_{2}x+\mu\Lambda_{1}\eta-\mu\Lambda_{2}\eta)\\
		& +x^{\mathrm{T}}P(\Lambda_{3}x+\mu\Lambda_{4}+\Lambda_{5})\\
		& +T||\dot{x}(t)||^{2}-\int_{t-T}^{t}||\dot{x}(s)||^{2}ds\\
		= & -\mu||x||^{2}+x^{\mathrm{T}}(\mu P\Lambda_{2}+P\Lambda_{3})x\\
		& +x^{\mathrm{T}}P(\mu\Lambda_{1}\eta-\mu\Lambda_{2}\eta)+x^{\mathrm{T}}P(\mu\Lambda_{4}+\Lambda_{5})\\
		& +T||\dot{x}(t)||^{2}-\int_{t-T}^{t}||\dot{x}(s)||^{2}ds.
		\end{align*}
		Noting that $\Lambda_{j}(j=2,3,4,5)\in\mathsf{E}(\gamma)$ and using
		Young's inequality, we have 
		\begin{align}
		\dot{V}\leq & -\mu||x||^{2}+\overline{\mu}c_{1}\mathrm{e}^{-\gamma t}||x||^{2}\nonumber \\
		& +\frac{\mu}{4}||x||^{2}+c_{2}\mu||\eta||^{2}\nonumber \\
		& +\frac{\mu}{4}||x||^{2}+\overline{\mu}c_{3}\mathrm{e}^{-2\gamma t}\nonumber \\
		& +T||\dot{x}(t)||^{2}-\int_{t-T}^{t}||\dot{x}(s)||^{2}ds\label{eq:29-4}
		\end{align}
		where $\overline{\mu}=\sqrt{\mu^{2}+1}$ and $c_{1},c_{2},c_{3}$
		are some positive constants.
		
		Meanwhile, using (\ref{eq:27-4}) for $||\dot{x}(t)||^{2}$ and Young's
		inequality, 
		\begin{equation}
		||\dot{x}(t)||^{2}\leq c_{4}\overline{\mu}^{2}||x||^{2}+c_{5}\mu^{2}||\eta||^{2}+c_{6}\overline{\mu}^{2}\mathrm{e}^{-2\gamma t}\label{eq:29-3}
		\end{equation}
		where $c_{4},c_{5},c_{6}$ are some positive constants.
		
		For $\eta$, by Jensen's inequality \cite{key-26}, we have
		\[
		\eta_{i}^{2}(t)=\left(\int_{t-d_{i}(t)}^{t}\dot{x}_{i}(s)ds\right)^{2}\leq T\int_{t-T}^{t}||\dot{x}_{i}(s)||^{2}ds,
		\]
		then
		\begin{equation}
		||\eta||^{2}\leq T\int_{t-T}^{t}||\dot{x}(s)||^{2}ds.\label{eq:12-2}
		\end{equation}
		Substituting (\ref{eq:29-3}) and (\ref{eq:12-2}) into (\ref{eq:29-4}),
		we get
		\begin{align*}
		\dot{V}\leq & -\left(\frac{\mu}{2}-\overline{\mu}c_{1}\mathrm{e}^{-\gamma t}-Tc_{4}\overline{\mu}^{2}\right)||x||^{2}\\
		& -(1-Tc_{2}\mu-Tc_{5}\mu^{2})\int_{t-T}^{t}||\dot{x}(s)||^{2}ds\\
		& +\overline{\mu}c_{3}\mathrm{e}^{-2\gamma t}+Tc_{6}\overline{\mu}^{2}\mathrm{e}^{-2\gamma t}.
		\end{align*}
		Then, for a positive constant $\gamma$, we have
		\begin{align}
		\dot{V}+\gamma V\leq & -\left(\frac{\mu}{2}-\overline{\mu}c_{1}\mathrm{e}^{-\gamma t}-Tc_{4}\overline{\mu}^{2}-\gamma||P||\right)||x||^{2}\nonumber \\
		& -(1-Tc_{2}\mu-Tc_{5}\mu^{2}-\gamma T)\int_{t-T}^{t}||\dot{x}(s)||^{2}ds\nonumber \\
		& +\overline{\mu}c_{3}\mathrm{e}^{-2\gamma t}+Tc_{6}\overline{\mu}^{2}\mathrm{e}^{-2\gamma t}.\label{eq:29-2}
		\end{align}
		
		Next, we will show $V$ does not exhibit finite time escape. From
		(\ref{eq:29-2}), we have
		\begin{align*}
		\dot{V}\leq & \alpha V+\beta
		\end{align*}
		where $\alpha,\beta$ are positive constants.
		
		This means 
		\begin{align*}
		V\leq V(0)\mathrm{e}^{\alpha t}-\frac{\beta}{\alpha}(1- & \mathrm{e}^{\alpha t}).
		\end{align*}
		Therefore, $V$ is bounded on finite time interval.
		
		Moreover, on a finite time interval $[0,t_{0})\subset[0,+\infty)$,
		we have
		\begin{align}
		V\leq & V(0)\mathrm{e}^{(\alpha+\gamma)t}\mathrm{e}^{-\gamma t}-\frac{\beta}{\alpha}(1-\mathrm{e}^{\alpha t})\mathrm{e}^{\gamma t}\mathrm{e}^{-\gamma t}\nonumber \\
		\leq & \max\{V(0)\mathrm{e}^{(\alpha+\gamma)t_{0}},\frac{\beta}{\alpha}(1-\mathrm{e}^{\alpha t_{0}})\mathrm{e}^{\gamma t_{0}}\}\mathrm{e}^{-\gamma t}\nonumber \\
		\leq & c_{7}\mathrm{e}^{-\gamma t}\label{eq:33-1}
		\end{align}
		where $c_{7}$ is a positive constant.
		
		On the other hand, for (\ref{eq:29-2}), there exists a finite time
		instant $t_{0}$, $\mu$ and $T$ such that
		\[
		\frac{\mu}{2}-\overline{\mu}c_{1}\mathrm{e}^{-\gamma t_{0}}-Tc_{4}\overline{\mu}^{2}-\gamma||P||>0,
		\]
		\[
		1-Tc_{2}\mu-Tc_{5}\mu^{2}-\gamma T>0.
		\]
		Therefore,
		\begin{align}
		\dot{V}\leq & -\gamma V+c_{8}\mathrm{e}^{-2\gamma t}\label{eq:43-1}
		\end{align}
		for $\forall t\in[t_{0},+\infty)$ where $c_{8}$ is a positive constant.
		
		Then, by solving the above inequality,
		\begin{align}
		V\leq & V(t_{0})\mathrm{e}^{-\gamma(t-t_{0})}-\frac{c_{8}}{\gamma}\mathrm{e}^{-2\gamma(t-t_{0})}+\frac{c_{8}}{\gamma}\mathrm{e}^{-\gamma t_{0}}\mathrm{e}^{-\gamma t}\nonumber \\
		\leq & \max\{V(t_{0}),\frac{c_{8}}{\gamma}\}\mathrm{e}^{-\gamma t}\leq c_{9}\mathrm{e}^{-\gamma t}\label{eq:33}
		\end{align}
		for $\forall t\in[t_{0},+\infty)$ where $c_{9}$ is a positive constant. 
		
		Then by combining (\ref{eq:33-1}) and (\ref{eq:33}), we can complete
		the proof.
	\end{IEEEproof}
	\begin{lem}
		\label{lem:Consider-the-following-1}\textcolor{black}{Consider the
			system (\ref{eq:26-3}) in a special form by letting $\Lambda_{2}=\Lambda_{3}=\Lambda_{5}=0$.
			That is
			\begin{align}
			\dot{x}= & -\mu\Lambda_{1}\boldsymbol{x}_{d}+\mu\Lambda_{4}\label{eq:26-3-1}
			\end{align}
			where $-\Lambda_{1}$ is a Hurwitz matrix, $\Lambda_{4}\in\mathsf{E}(\gamma)$
			with a positive constant $\gamma$. If $\mu,T$ satisfy
			\begin{align}
			T<\mathrm{min} & \left\{ \frac{1}{6\mu||\Lambda_{1}||^{2}}-\frac{\gamma||P||}{3\mu^{2}||\Lambda_{1}||^{2}}\right.,\nonumber \\
			& \left.\frac{1}{\mu||P\Lambda_{1}||+3\mu^{2}||P\Lambda_{1}||+\gamma}\right\} \label{eq:45}
			\end{align}
			where $P$ is a positive definite matrix such that $P\Lambda_{1}+\Lambda_{1}^{\mathrm{T}}P=2I$,
			then $x\in\mathsf{E}(\gamma)$.}
	\end{lem}
	\begin{IEEEproof}
		The proof follows the line of the proof of 1). Under the assumption
		that $\Lambda_{2}=\Lambda_{3}=\Lambda_{5}=0$, (\ref{eq:26-3}) can
		be written as:
		\begin{align}
		\dot{x}= & -\mu\Lambda_{1}x+\mu\Lambda_{1}\eta+\mu\Lambda_{4}\label{eq:44-1}
		\end{align}
		where $\eta(t)=\mathrm{col}(\eta_{1}(t),\eta_{2}(t),...,\eta_{N}(t))$
		with $\eta_{i}(t)=\int_{t-d_{i}(t)}^{t}\dot{x}_{i}(s)ds$.
		
		Consider a Lyapunov-Krasovskii function in the form of (\ref{eq:13-1}).
		By (\ref{eq:44-1}), we have
		\begin{align*}
		\dot{V}\leq & x^{\mathrm{T}}P(-\mu\Lambda_{1}x+\mu\Lambda_{1}\eta+\mu\Lambda_{4})\\
		& +T||\dot{x}(t)||^{2}-\int_{t-T}^{t}||\dot{x}(s)||^{2}ds.
		\end{align*}
		By Young's inequality, we get
		\begin{align}
		\dot{V}\leq & -\mu||x||^{2}\nonumber \\
		& +\frac{\mu}{4}||x||^{2}+\mu||P\Lambda_{1}||\cdot||\eta||^{2}\nonumber \\
		& +\frac{\mu}{4}||x||^{2}+\mu c_{10}\mathrm{e}^{-2\gamma t}\nonumber \\
		& +T||\dot{x}(t)||^{2}-\int_{t-T}^{t}||\dot{x}(s)||^{2}ds\label{eq:29-4-1}
		\end{align}
		where $c_{10}$ is a positive constant.
		
		Meanwhile, using (\ref{eq:44-1}) for $||\dot{x}(t)||^{2}$ and by
		Young's inequality
		\begin{equation}
		||\dot{x}(t)||^{2}\leq3\mu^{2}||\Lambda_{1}||^{2}||x||^{2}+3\mu^{2}||\Lambda_{1}||^{2}||\eta||^{2}+c_{11}\mu^{2}\mathrm{e}^{-2\gamma t}\label{eq:29-3-1}
		\end{equation}
		where $c_{11}$ is a positive constant.
		
		Substituting (\ref{eq:29-3-1}) into (\ref{eq:29-4-1}),
		\begin{align*}
		\dot{V}\leq & -\left(\frac{\mu}{2}-3T\mu^{2}||\Lambda_{1}||^{2}\right)||x||^{2}\\
		& -(1-T\mu||P\Lambda_{1}||-3T\mu^{2}||\Lambda_{1}||^{2})\int_{t-T}^{t}||\dot{x}(s)||^{2}ds\\
		& +\mu c_{10}\mathrm{e}^{-2\gamma t}+Tc_{11}\mu^{2}\mathrm{e}^{-2\gamma t}.
		\end{align*}
		Then, for a positive constant $\gamma$, we have
		\begin{align}
		& \dot{V}+\gamma V\nonumber \\
		\leq & -\left(\frac{\mu}{2}-3T\mu^{2}||\Lambda_{1}||^{2}-\gamma||P||\right)||x||^{2}\nonumber \\
		& -(1-T\mu||P\Lambda_{1}||-3T\mu^{2}||\Lambda_{1}||^{2}-\gamma T)\int_{t-T}^{t}||\dot{x}(s)||^{2}ds\nonumber \\
		& +\mu c_{10}\mathrm{e}^{-2\gamma t}+Tc_{11}\mu^{2}\mathrm{e}^{-2\gamma t}.\label{eq:29-2-2}
		\end{align}
		If the following inequality holds
		\begin{equation}
		\frac{\mu}{2}-3T\mu^{2}||\Lambda_{1}||^{2}-\gamma||P||>0,\label{eq:49}
		\end{equation}
		\begin{equation}
		1-T\mu||P\Lambda_{1}||-3T\mu^{2}||\Lambda_{1}||^{2}-\gamma T>0,\label{eq:50}
		\end{equation}
		then we have
		\begin{align*}
		\dot{V}\leq & -\gamma V+\mu c_{10}\mathrm{e}^{-2\gamma t}+Tc_{11}\mu^{2}\mathrm{e}^{-2\gamma t}.
		\end{align*}
		By solving the above inequality, we can show $x\in\mathsf{E}(\gamma)$.
		Finally, note that (\ref{eq:49})-(\ref{eq:50}) are equivalent to
		(\ref{eq:45}). This completes the proof.
	\end{IEEEproof}
	
	\subsection{Proof of Theorems \ref{thm:1-1} and \ref{thm:1}}
	\begin{IEEEproof}
		We will first prove Theorem \ref{thm:1}. The proof is divided into
		the following two steps.
		
		\textit{Step 1. }Show\textit{ }$\tilde{A}_{i}\triangleq\hat{A}_{i}-A(i=1,2,...,N)$
		converges to zero exponentially.
		
		Note that (\ref{eq:5-1-1}) can be transformed into
		\begin{align}
		\dot{\tilde{A}} & =-\mu_{1}(\mathcal{H}\varotimes I)(\hat{\boldsymbol{A}}(\overline{t}_{l})-\overline{A})\nonumber \\
		& =-\mu_{1}(\mathcal{H}\varotimes I)(\hat{\boldsymbol{A}}(\overline{t}_{l})-\hat{\boldsymbol{A}}(t_{k})+\tilde{\boldsymbol{A}}(t_{k}))\label{eq:31-1}
		\end{align}
		where $\overline{A}=\mathrm{col}(A,A,...,A)$, $\hat{A}=\mathrm{col}(\hat{A}_{1},\hat{A}_{2},...,\hat{A}_{N})$,
		$\tilde{A}=\hat{A}-\overline{A}$. $\hat{\boldsymbol{A}}(\overline{t}_{l})=\mathrm{col}(\hat{A}_{1}(\overline{t}_{l'}^{1}),\hat{A}_{2}(\overline{t}_{l'}^{2}),...,\hat{A}_{N}(\overline{t}_{l'}^{N}))$,
		$\hat{\boldsymbol{A}}(t_{k})=\mathrm{col}(\hat{A}_{1}(t_{k'}^{1}),\hat{A}_{2}(t_{k'}^{2}),...,\hat{A}_{N}(t_{k'}^{N}))$,
		$\tilde{\boldsymbol{A}}(t_{k})=\hat{\boldsymbol{A}}(t_{k})-\overline{A}$.
		
		Let $\overline{\alpha}=\mathrm{vec}(\overline{A})$, $\hat{\alpha}=\mathrm{vec}(\hat{A})$,
		$\tilde{\alpha}=\mathrm{vec}(\tilde{A})$, $\hat{\boldsymbol{\alpha}}(\overline{t}_{l})=\mathrm{vec}(\hat{\boldsymbol{A}}(\overline{t}_{l}))$,
		$\hat{\boldsymbol{\alpha}}(t_{k})=\mathrm{vec}(\hat{\boldsymbol{A}}(t_{k}))$
		and $\tilde{\boldsymbol{\alpha}}(t_{k})=\mathrm{vec}(\tilde{\boldsymbol{A}}(t_{k}))$,
		(\ref{eq:31-1}) becomes
		\begin{align*}
		\dot{\tilde{\alpha}} & =-\mu_{1}(I\varotimes\mathcal{H}\varotimes I)(\hat{\boldsymbol{\alpha}}(\overline{t}_{l}^{i})-\hat{\boldsymbol{\alpha}}(t_{k})+\tilde{\boldsymbol{\alpha}}(t_{k})).
		\end{align*}
		
		It follows that
		\begin{align}
		\dot{\tilde{\alpha}} & =-\mu_{1}(I\varotimes\mathcal{H}\varotimes I)\tilde{\boldsymbol{\alpha}}(t_{k})+\mu_{1}\Lambda_{A}\label{eq:36-1}
		\end{align}
		\textcolor{black}{where $\Lambda_{A}=(I\varotimes\mathcal{H}\varotimes I)(\hat{\boldsymbol{\alpha}}(\overline{t}_{l})-\hat{\boldsymbol{\alpha}}(t_{k}))$.}
		
		\textcolor{black}{From the event-triggered condition (\ref{eq:26}),
			we know $\Lambda_{A}\in\mathsf{E}(\gamma_{A})$. Then, let $d_{i}(t)=t-t_{k}^{i}$
			in (\ref{eq:36-1}) and use Lemma \ref{lem:Consider-the-following}
			in Appendix A, we can show $\tilde{\alpha}\in\mathsf{E}(\gamma_{A})$,
			$i.e.$, $\tilde{A}_{i}\in\mathsf{E}(\gamma_{A})(i=1,2,...,N)$.}
		
		\textit{Step 2. }Show\textit{ }$\tilde{\nu}_{i}\triangleq\hat{\nu}_{i}-\nu(i=1,2,...,N)$
		converges to zero exponentially.
		
		Let
		\begin{align}
		z_{i}(t) & =\mathrm{e}^{-At}\hat{\nu}_{i}(t)(i=0,1,...,n)\label{eq:9-2-1}
		\end{align}
		where $z_{0}(t)=\mathrm{e}^{-At}\hat{\nu}_{0}(t)=\mathrm{e}^{-At}\nu(t)=\nu(0).$
		
		Then, (\ref{eq:6-1-1}) can be expressed as:
		\begin{align*}
		\dot{z}_{i}= & -\mathrm{e}^{-At}A\hat{\nu}_{i}(t)+\mathrm{e}^{-At}\hat{A}_{i}\hat{\nu}_{i}(t)\\
		& +\mathrm{e}^{-At}\mu_{2}\sum_{j=0}^{N}a_{ij}(\overline{\nu}_{j}(t,\overline{t}_{l'}^{j})-\overline{\nu}_{i}(t,\overline{t}_{l}^{i})).
		\end{align*}
		The above inequality can be written as:
		\begin{align*}
		\dot{z}_{i}= & \Delta_{i}^{1}+\mu_{2}\sum_{j=0}^{N}a_{ij}\left(\Delta_{j}^{2}-\Delta_{i}^{2}\right)\\
		& +\mu_{2}\sum_{j=0}^{N}a_{ij}\left(\Delta_{j}^{3}-\Delta_{i}^{3}\right)\\
		& +\mu_{2}\sum_{j=0}^{N}a_{ij}(z_{j}(t_{k'}^{j})-z_{i}(t_{k}^{i}))
		\end{align*}
		where 
		\begin{equation}
		\Delta_{i}^{1}=\mathrm{e}^{-At}\hat{A}_{i}\hat{\nu}_{i}(t)-\mathrm{e}^{-At}A\hat{\nu}_{i}(t),
		\end{equation}
		\begin{equation}
		\Delta_{i}^{2}=\mathrm{e}^{-At}\left(\mathrm{e}^{\hat{A}_{i}(\overline{t}_{l}^{i})(t-\overline{t}_{l}^{i})}\hat{\nu}_{i}(\overline{t}_{l}^{i})-\mathrm{e}^{\hat{A}_{i}(\overline{t}_{l}^{i})(t-t_{k}^{i})}\hat{\nu}_{i}(t_{k}^{i})\right),\label{eq:39}
		\end{equation}
		\begin{equation}
		\Delta_{i}^{3}=\mathrm{e}^{-At}\left(\mathrm{e}^{\hat{A}_{i}(\overline{t}_{l}^{i})(t-t_{k}^{i})}\hat{\nu}_{i}(t_{k}^{i})-\mathrm{e}^{A(t-t_{k}^{i})}\hat{\nu}_{i}(t_{k}^{i})\right).\label{eq:40}
		\end{equation}
		By considering agent $i=1,2,...,N$, we have
		\begin{align}
		\dot{\tilde{z}}= & -\mu_{2}(\mathcal{H}\varotimes I)\tilde{\boldsymbol{z}}(t_{k})\nonumber \\
		& +\Delta^{1}-\mu_{2}(\mathcal{H}\varotimes I)\Delta^{2}-\mu_{2}(\mathcal{H}\varotimes I)\Delta^{3}\label{eq:41-1}
		\end{align}
		where $\tilde{\boldsymbol{z}}(t_{k})=\mathrm{col}(z_{1}(t_{k'}^{1}),z_{2}(t_{k'}^{2}),...,z_{N}(t_{k'}^{N}))$,
		$\Delta^{1}=\mathrm{col}(\Delta_{1}^{1},\Delta_{2}^{1},...,\Delta_{N}^{1})$,
		$\Delta^{2}=\mathrm{col}(\Delta_{1}^{2},\Delta_{2}^{2},...,\Delta_{N}^{2})$,
		$\Delta^{3}=\mathrm{col}(\Delta_{1}^{3},\Delta_{2}^{3},...,\Delta_{N}^{3})$. 
		
		In the following, we will have an analysis on $\Delta^{1},\Delta^{2}$
		and $\Delta^{3}$.
		
		For $\Delta_{i}^{1}$ in $\Delta^{1}$, we have
		\begin{align*}
		\Delta_{i}^{1} & =\mathrm{e}^{-At}\hat{A}_{i}\hat{\nu}_{i}(t)-\mathrm{e}^{-At}A\hat{\nu}_{i}(t)\\
		& =\mathrm{e}^{-At}\tilde{A}_{i}\mathrm{e}^{At}\tilde{z}_{i}+\mathrm{e}^{-At}\tilde{A}_{i}\mathrm{e}^{At}z_{0}.
		\end{align*}
		This implies that
		\begin{align}
		\Delta^{1} & =\Lambda_{1}\tilde{z}+\Lambda_{1}\bar{z}\label{eq:42-1}
		\end{align}
		where $\Lambda_{1}=\mathrm{diag}(\mathrm{e}^{-At}\tilde{A}_{1}\mathrm{e}^{At},\mathrm{e}^{-At}\tilde{A}_{2}\mathrm{e}^{At},...,\mathrm{e}^{-At}\tilde{A}_{N}\mathrm{e}^{At})$,
		$\bar{z}=\mathrm{col}(z_{0},z_{0},...,z_{0})$.
		
		For $i=1,2,...,N$, using the result in Step 1 and Lemma \ref{lem:1)-(Gronwall's-inequality)}-1),
		we get
		\[
		||\mathrm{e}^{-At}\tilde{A}_{i}\mathrm{e}^{At}||\leq||\mathrm{e}^{-At}||\cdot||\tilde{A}_{i}||\cdot||\mathrm{e}^{-At}||=||\tilde{A}_{i}||\in\mathsf{E}(\gamma_{A}).
		\]
		This implies that $||\Lambda_{1}||\in\mathsf{E}(\gamma_{A})$ and
		$||\Lambda_{1}\bar{z}||\in\mathsf{E}(\gamma_{A})$.
		
		For $\Delta_{i}^{2}$ in $\Delta^{2}$, according to (\ref{eq:39}),
		we have
		\begin{align*}
		\Delta_{i}^{2} & =\mathrm{e}^{-At}\mathrm{e}^{\hat{A}_{i}(\overline{t}_{l}^{i})(t-t_{k}^{i})}\left(\mathrm{e}^{\hat{A}_{i}(\overline{t}_{l}^{i})(t_{k}^{i}-\overline{t}_{l}^{i})}\hat{\nu}_{i}(\overline{t}_{l}^{i})-\hat{\nu}_{i}(t_{k}^{i})\right).
		\end{align*}
		Note that $||\mathrm{e}^{-At}\mathrm{e}^{\hat{A}_{i}(\overline{t}_{l}^{i})(t-t_{k}^{i})}||$
		are bounded by Lemma \ref{lem:1)-(Gronwall's-inequality)}-1). According
		to (\ref{eq:26}), we get $||\Delta_{i}^{2}||\in\mathsf{E}(\gamma_{\nu}).$
		Thus, $||(\mathcal{H}\varotimes I)\Delta^{2}||\in\mathsf{E}(\gamma_{\nu}).$
		
		For $\Delta_{i}^{3}$ in $\Delta^{3}$, according to (\ref{eq:40}),
		we have
		\begin{align*}
		\Delta_{i}^{3} & =\mathrm{e}^{-At}\left(\mathrm{e}^{\hat{A}_{i}(t-t_{k}^{i})}-\mathrm{e}^{A(t-t_{k}^{i})}\right)\mathrm{e}^{At_{k}^{i}}z_{i}(t_{k}^{i}).
		\end{align*}
		This indicates that 
		\begin{equation}
		\Delta^{3}=\Lambda_{2}\tilde{\boldsymbol{z}}(t_{k})+\Lambda_{2}\bar{z}\label{eq:43}
		\end{equation}
		where $\bar{z}=\mathrm{col}(z_{0},z_{0},...,z_{0})$, $\Lambda_{2}=\mathrm{diag}(\mathrm{e}^{-At}(\mathrm{e}^{\hat{A}_{1}(t-t_{k'}^{1})}-\mathrm{e}^{A(t-t_{k'}^{1})})\mathrm{e}^{At_{k'}^{1}},...,\mathrm{e}^{-At}(\mathrm{e}^{\hat{A}_{N}(t-t_{k'}^{N})}-\mathrm{e}^{A(t-t_{k'}^{N})})\mathrm{e}^{At_{k'}^{N}})$.
		
		Note that for any entry in $\Lambda_{2}$, using Lemma \ref{lem:1)-(Gronwall's-inequality)}-2),
		we have
		\begin{align*}
		& ||\mathrm{e}^{-At}(\mathrm{e}^{\hat{A}_{i}(t-t_{k}^{i})}-\mathrm{e}^{A(t-t_{k}^{i})})\mathrm{e}^{At_{k}^{i}}||\\
		\leq & ||\mathrm{e}^{-At}||\cdot||\mathrm{e}^{\hat{A}_{i}(t-t_{k}^{i})}-\mathrm{e}^{A(t-t_{k}^{i})}||\cdot||\mathrm{e}^{At_{k}^{i}}||\\
		\leq & ||\hat{A}_{i}-A||\mathrm{e}^{||(\hat{A}_{i}-A)(t-t_{k}^{i})||+||A(t-t_{k}^{i})||}\in\mathsf{E}(\gamma_{A}).
		\end{align*}
		This implies that $||\Lambda_{2}||\in\mathsf{E}(\gamma_{A})$ and
		$||\Lambda_{2}\bar{z}||\in\mathsf{E}(\gamma_{A})$.
		
		Based on (\ref{eq:42-1}) and (\ref{eq:43}), (\ref{eq:41-1}) becomes
		\begin{align}
		\dot{\tilde{z}}= & -\mu_{2}(\mathcal{H}\varotimes I)\tilde{\boldsymbol{z}}(t_{k})\nonumber \\
		& +\Lambda_{1}\tilde{z}+\Lambda_{1}\bar{z}-\mu_{2}(\mathcal{H}\varotimes I)\Delta^{2}\nonumber \\
		& -\mu_{2}(\mathcal{H}\varotimes I)(\Lambda_{2}\tilde{\boldsymbol{z}}(t_{k})+\Lambda_{2}\bar{z})\nonumber \\
		= & -\mu_{2}(\mathcal{H}\varotimes I)\tilde{\boldsymbol{z}}(t_{k})-\mu_{2}(\mathcal{H}\varotimes I)\Lambda_{2}\tilde{\boldsymbol{z}}(t_{k})+\Lambda_{1}\tilde{z}\nonumber \\
		& -\mu_{2}(\mathcal{H}\varotimes I)\Delta^{2}-\mu_{2}(\mathcal{H}\varotimes I)\Lambda_{2}\bar{z}+\Lambda_{1}\bar{z}.\label{eq:44}
		\end{align}
		Let $\mathcal{H}\varotimes I=\overline{\Lambda}_{1}$, $-\mu_{2}(\mathcal{H}\varotimes I)\Lambda_{2}=\overline{\Lambda}_{2}$,
		$\Lambda_{1}=\overline{\Lambda}_{3}$, $-(\mathcal{H}\varotimes I)\Delta^{2}-(\mathcal{H}\varotimes I)\Lambda_{2}\bar{z}=\overline{\Lambda}_{4}$,
		$\Lambda_{1}\bar{z}=\overline{\Lambda}_{5}$. 
		
		Then, (\ref{eq:44}) is expressed as 
		\begin{align*}
		\dot{\tilde{z}}= & -\mu_{2}\overline{\Lambda}_{1}\tilde{\boldsymbol{z}}(t_{k})+\mu_{2}\overline{\Lambda}_{2}\tilde{\boldsymbol{z}}(t_{k})+\overline{\Lambda}_{3}\tilde{z}\\
		& +\mu_{2}\overline{\Lambda}_{4}+\overline{\Lambda}_{5}
		\end{align*}
		where $\overline{\Lambda}_{2},\overline{\Lambda}_{3},\overline{\Lambda}_{4},\overline{\Lambda}_{5}\in\mathsf{E}(\mathrm{min}(\gamma_{A},\gamma_{\nu})).$
		
		Using Lemma \ref{lem:Consider-the-following} in Appendix A, we can
		show $\tilde{z}\in\mathsf{E}(\mathrm{min}(\gamma_{A},\gamma_{\nu}))$.
		Note that
		\begin{align*}
		||\tilde{\nu}_{i}||=||\mathrm{e}^{-At}\tilde{\nu}_{i}||\leq||\tilde{z}||\in\mathsf{E}(\mathrm{min}(\gamma_{A},\gamma_{\nu}))
		\end{align*}
		for $i=1,2,...,N$. Therefore, Theorem \ref{thm:1} is proved.\medskip{}
		
		\textcolor{black}{Next, we will prove Theorem \ref{thm:1-1}. The proof
			follows the line of Step 2 by using the real value of $A$ instead
			of $\hat{A}_{i}$. In this case, (\ref{eq:41-1}) becomes
			\begin{align*}
			\dot{\tilde{z}}= & -\mu_{2}(\mathcal{H}\varotimes I)\tilde{\boldsymbol{z}}(t_{k})+\mu_{2}\overline{\Lambda}_{4}
			\end{align*}
			where $\overline{\Lambda}_{4}\in\mathsf{E}(\gamma_{\nu}).$}
		
		\textcolor{black}{Then, by Lemma \ref{lem:Consider-the-following-1},
			we can show $||\tilde{\nu}_{i}||=||\mathrm{e}^{-At}\tilde{\nu}_{i}||\leq||\tilde{z}||\in\mathsf{E}(\gamma_{\nu})$
			if $T$ satisfies (\ref{eq:45-1}). This completes the proof.}
	\end{IEEEproof}
	
	\subsection{Proof of Proposition \ref{prop:1}}
	\begin{IEEEproof}
		First, from (\ref{eq:22-3}), we know 
		\begin{align*}
		\xi_{i2} & =\overline{f}_{i1}(\overline{z}_{i},\tilde{\eta}_{i1},\overline{x}_{i1},\nu,w)+b_{i1}(w)\overline{x}_{i2}\\
		& \triangleq\overline{\xi}_{i2}(\overline{z}_{i},\tilde{\eta}_{i1},\overline{x}_{i1},\overline{x}_{i2},\nu,w).
		\end{align*}
		This shows (\ref{eq:11-2}) holds with $j=2$. Meanwhile, we have
		\begin{align*}
		\overline{x}_{i2} & =(\xi_{i2}-\overline{f}_{i1}(\overline{z}_{i},\tilde{\eta}_{i1},\xi_{i1},\nu,w))/b_{i1}(w)\\
		& \triangleq\chi_{i2}(\overline{z}_{i},\tilde{\eta}_{i1},\xi_{i1},\xi_{i2},\nu,w).
		\end{align*}
		This shows (\ref{eq:12-1}) holds with $j=2$.
		
		For $j=3$, by (\ref{eq:22-3}) and (\ref{eq:6}), we know 
		\begin{align*}
		\xi_{i3}\triangleq\dot{\xi}_{i2}= & \frac{\partial\overline{\xi}_{i2}}{\partial\overline{z}_{i}}\overline{f}_{i0}+\frac{\partial\overline{\xi}_{i2}}{\partial\tilde{\eta}_{i1}}(M_{ij}\tilde{\eta}_{ij}+g_{ij})\\
		& +\frac{\partial\overline{\xi}_{i2}}{\partial\overline{x}_{i1}}\left(\overline{f}_{i1}+b_{i1}(w)\overline{x}_{i2}\right)\\
		& +\frac{\partial\overline{\xi}_{i2}}{\partial\overline{x}_{i2}}\left(\overline{f}_{i2}+b_{i2}(w)\overline{x}_{i3}\right)+\frac{\partial\overline{\xi}_{i2}}{\partial\nu}A\nu\\
		\triangleq & \overline{\xi}_{i3}(\overline{z}_{i},\tilde{\eta}_{i1},\tilde{\eta}_{i2},\overline{x}_{i1},\overline{x}_{i2},\overline{x}_{i3},\nu,w).
		\end{align*}
		This shows (\ref{eq:11-2}) holds with $j=3$. Similarly, we can show
		(\ref{eq:12-1}) holds with $j\geq4$. This completes the proof.
	\end{IEEEproof}
	
	\subsection{Proof of Proposition \ref{prop:2}}
	\begin{IEEEproof}
		For $j=1$, we have 
		\[
		\dot{\alpha}_{i1}=-Q_{i1}\dot{\zeta}_{i1}=-Q_{i1}(\zeta_{i2}-Q_{i1}\zeta_{i1})\leq\vartheta_{i1}(Q_{i1})(|\zeta_{i1}|+|\zeta_{i2}|).
		\]
		
		For $j=2$, using the above inequality, we have 
		\begin{align*}
		\dot{\alpha}_{i2} & =-Q_{i2}\dot{\zeta}_{i2}=-Q_{i2}(\zeta_{i3}+\alpha_{i2}-\dot{\alpha}_{i1})\\
		& =-Q_{i2}(\zeta_{i3}-Q_{i2}\zeta_{i2}-\dot{\alpha}_{i1})\\
		& \leq\vartheta_{i2}(Q_{i1},Q_{i2})(|\zeta_{i1}|+|\zeta_{i2}|+|\zeta_{i3}|).
		\end{align*}
		
		By repeating the above procedures for $j=3,4,...,n$, we can complete
		the proof.
	\end{IEEEproof}
	
	\subsection{Proof of Lemma \ref{lem:For-system-(),}}
	\begin{IEEEproof}
		The proof is divided into the following steps. We will analyze each
		term in the Lyapunov function (\ref{eq:24}).
		
		\textit{Step 1)}. \textit{Analysis of $V_{i0}(\overline{z}_{i})$.}
		
		According to Lemma 11.1 in \cite{key-3}, we know when $X_{i}\in\Omega_{R}$,
		there exists a positive constant gain $\overline{\mu}_{i0}(R)$ related
		with $R$ such that
		\begin{align*}
		& \left\Vert \frac{\partial V_{i0}}{\partial\overline{z}_{i}}\right\Vert \cdot||\overline{f}_{i0}(\overline{z}_{i},\zeta_{1i},\nu,w)-\overline{f}_{i0}(\overline{z}_{i},0,\nu,w)||\\
		\leq & \overline{\mu}_{i0}(R)||\overline{z}_{i}||\cdot|\zeta_{1i}|.
		\end{align*}
		
		Then, using Assumption \ref{fact:=00005B=00005D-Assume-there} and
		Young's inequality, the derivative of $V_{i0}$ can be computed as:
		\begin{align*}
		\dot{V}_{i0}(\overline{z}_{i})= & \frac{\partial V_{i0}}{\partial\overline{z}_{i}}\overline{f}_{i0}(\overline{z}_{i},\zeta_{1i},\nu,w)\\
		= & \frac{\partial V_{i0}}{\partial\overline{z}_{i}}\overline{f}_{i0}(\overline{z}_{i},0,\nu,w)\\
		& +\frac{\partial V_{i0}}{\partial\overline{z}_{i}}(\overline{f}_{i0}(\overline{z}_{i},\zeta_{1i},\nu,w)-\overline{f}_{i0}(\overline{z}_{i},0,\nu,w))\\
		\leq & -\gamma_{i0}||\overline{z}_{i}||^{2}+\overline{\mu}_{i0}(R)||\overline{z}_{i}||\cdot|\zeta_{1i}|\\
		\leq & -\frac{\gamma_{i0}|}{2}|\overline{z}_{i}||^{2}+\mu_{i0}(R)\zeta_{1i}^{2}
		\end{align*}
		where $\mu_{i0}(R)$ is a positive constant related with $R$. 
		
		Then, let $V_{i}^{z}=\frac{V_{i0}(\overline{z}_{i})}{L_{i0}}$, we
		have
		\begin{align*}
		\dot{V}_{i}^{z}(\overline{z}_{i})\leq & -\frac{\gamma_{i0}}{2L_{i0}}||\overline{z}_{i}||^{2}+\frac{\mu_{i0}(R)}{L_{i0}}\zeta_{1i}^{2}.
		\end{align*}
		
		\textit{Step 2)}. \textit{Analysis of} $\frac{\tilde{\eta}_{ij}^{\mathrm{T}}P_{ij}\tilde{\eta}_{ij}}{L_{ij}}$.
		
		Consider the following Lyapunov function $V_{ij}^{\eta}=\frac{\tilde{\eta}_{ij}^{\mathrm{T}}P_{ij}\tilde{\eta}_{ij}}{L_{ij}}.$
		Then using Lemma 11.1 in \cite{key-3} and the transformed system
		(\ref{eq:22-1}), the derivative of $V_{ij}^{\eta}$ is 
		\begin{align}
		\dot{V}_{ij}^{\eta}= & \frac{1}{L_{ij}}\tilde{\eta}_{ij}^{\mathrm{T}}(P_{ij}M_{ij}+M_{ij}^{\mathrm{T}}P_{ij})\tilde{\eta}_{ij}\nonumber \\
		& +\frac{1}{L_{ij}}\tilde{\eta}_{ij}^{\mathrm{T}}P_{ij}\overline{h}_{ij}(\overline{z}_{i},\tilde{\eta}_{i1},...,\tilde{\eta}_{i,j-1},\zeta_{i1},...,\zeta_{ij},\nu,w)\nonumber \\
		\leq & -\frac{\beta_{ij}}{2L_{ij}}||\tilde{\eta}_{ij}||^{2}+\frac{\overline{\gamma}_{ij}(\overline{z}_{i})}{L_{ij}}||\overline{z}_{i}||^{2}+\sum_{k=1}^{j-1}\frac{\overline{\varrho}_{ijk}(\tilde{\eta}_{ik})}{L_{ij}}||\tilde{\eta}_{ik}||^{2}\nonumber \\
		& +\sum_{k=1}^{j}\frac{\overline{\mu}_{ijk}(\zeta_{ik})}{L_{ij}}\zeta_{ik}^{2}\label{eq:22}
		\end{align}
		where $\overline{\gamma}_{ij}(\overline{z}_{i}),\overline{\varrho}_{ijk}(\tilde{\eta}_{ik}),\overline{\mu}_{ijk}(\zeta_{ik})$
		are continuous functions.
		
		Note that by (\ref{eq:25}), for $\forall X_{i}\in\Omega_{R}$, 
		\[
		c_{P_{ij}}||\tilde{\eta}_{ij}||^{2}\leq\tilde{\eta}_{ij}^{T}P_{ij}\tilde{\eta}_{ij}\leq L_{ij}R^{2}
		\]
		where $c_{P_{ij}}>0$ denotes the minimum eigenvalue of matrix $P_{ij}$.
		
		Hence, (\ref{eq:22}) can be expressed as:
		\begin{align*}
		\dot{V}_{ij}^{\eta}\leq & -\frac{\beta_{ij}}{2L_{ij}}||\tilde{\eta}_{ij}||^{2}+\frac{\gamma_{ij}(R)}{L_{ij}}||\overline{z}_{i}||^{2}+\sum_{k=1}^{j-1}\frac{\varrho_{ijk}(RL_{ik})}{L_{ij}}||\tilde{\eta}_{ik}||^{2}\\
		& +\frac{\mu_{ij}(R)}{L_{ij}}\sum_{k=1}^{j}\zeta_{ik}^{2}
		\end{align*}
		where $\gamma_{ij}(R),\varrho_{ij}(RL_{ik}),\mu_{ij}(R)$ are positive
		gains related with $R$ and $L_{ik}$.
		
		\textit{Step 3)}. \textit{Analysis of} $V_{i}^{z}$ and $V_{ij}^{\eta}$.
		
		Consider the following Lyapunov function 
		\[
		W_{i}\triangleq V_{i}^{z}+\sum_{j=1}^{n}V_{ij}^{\eta}.
		\]
		We can obtain 
		\begin{align}
		\dot{W}_{i}= & \dot{V}_{i}^{z}(\overline{z}_{i})+\sum_{j=1}^{n}\dot{V}_{ij}^{\eta}\nonumber \\
		\leq & -\left(\frac{\gamma_{i0}}{2L_{i0}}-\sum_{j=1}^{n}\frac{\gamma_{ij}(R)}{L_{ij}}\right)||\overline{z}_{i}||^{2}\nonumber \\
		& -\sum_{j=1}^{n}\left(\frac{\beta_{ij}}{2L_{ij}}-\sum_{k=j+1}^{n}\frac{\varrho_{ijk}(RL_{ik})}{L_{ik}}\right)||\tilde{\eta}_{ij}||^{2}\nonumber \\
		& +\left(\sum_{j=0}^{n}\frac{\mu_{ij}(R)}{L_{ij}}\right)\zeta_{i1}^{2}+\sum_{k=2}^{n}\left(\sum_{j=k}^{n}\frac{\mu_{ij}(R)}{L_{ij}}\right)\zeta_{ik}^{2}.\label{eq:27-1}
		\end{align}
		Thus there exist sufficiently large scaling gains $L_{ij}$ and positive
		constants $\tilde{\gamma}_{i},\tilde{\varrho}_{i}$ such that
		\begin{align*}
		\frac{\gamma_{i0}}{2L_{i0}}-\sum_{j=1}^{n}\frac{\gamma_{ij}(R)}{L_{ij}} & \geq\tilde{\gamma}_{i},\\
		\frac{\beta_{ij}}{2L_{ij}}-\sum_{k=j+1}^{n}\frac{\varrho_{ijk}(RL_{ij})}{L_{ik}} & \geq\tilde{\varrho}_{i},j=1,...,n,\\
		\sum_{j=0}^{n}\frac{\mu_{ij}(R)}{L_{ij}} & \leq\frac{1}{4}.
		\end{align*}
		Hence, we have
		\[
		\dot{W}_{i}\leq-\tilde{\gamma}_{i}||\overline{z}_{i}||^{2}-\tilde{\varrho}_{i}\sum_{k=1}^{j}||\tilde{\eta}_{ij}||^{2}+\frac{1}{4}\sum_{j=1}^{n}\zeta_{ij}^{2}.
		\]
		
		\textit{Step 4)}. \textit{Analysis of} $\frac{1}{2}\zeta_{ij}^{2}$.
		
		Consider Lyapunov function $V_{ij}^{\zeta}=\frac{1}{2}\zeta_{ij}^{2}.$
		By (\ref{eq:22-1}), (\ref{eq:22-4}) and Proposition \ref{prop:2},
		we have for $j=1,...,n-1$,
		\begin{align*}
		\dot{V}_{ij}^{\zeta}= & \zeta_{ij}(\zeta_{i,j+1}-Q_{ij}\zeta_{ij}-\dot{\alpha}_{i,j-1})\\
		\leq & -Q_{ij}\zeta_{ij}^{2}+\zeta_{ij}\zeta_{i,j+1}\\
		& +|\zeta_{ij}|\vartheta_{ij}(Q_{i1},Q_{i2},...,Q_{i,j-1})(|\zeta_{i1}|+\cdots+|\zeta_{ij}|)\\
		\leq & -\left(Q_{ij}-\overline{\vartheta}_{ij}(Q_{i1},...,Q_{i,j-1})\right)\zeta_{ij}^{2}\\
		& +\frac{\zeta_{i,j+1}^{2}}{2}+\sum_{k=1}^{j-1}\zeta_{ik}^{2}
		\end{align*}
		where $\overline{\vartheta}_{ij}(Q_{i1},...,Q_{i,j-1})$ is a positive
		constant depending on $Q_{i1},...,Q_{i,j-1}$.
		
		Let
		\[
		Q_{ij}=\overline{\vartheta}_{ij}(Q_{i1},...,Q_{i,j-1})+n.
		\]
		We obtain
		\begin{align*}
		\dot{V}_{ij}^{\zeta}\leq & -n\zeta_{ij}^{2}+\frac{\zeta_{i,j+1}^{2}}{2}+\sum_{k=1}^{j-1}\zeta_{ik}^{2}(j=1,...,n-1).
		\end{align*}
		Then let $V_{i}^{\zeta}=\sum_{j=1}^{n}\frac{\zeta_{ij}^{2}}{2}$ and
		use (\ref{eq:22-1}). We have
		\begin{align*}
		\dot{V}_{i}^{\zeta}\leq & -\sum_{j=1}^{n-1}\zeta_{ij}^{2}+\zeta_{in}(\overline{\phi}_{i}-\dot{\alpha}_{i,n-1}+b_{in}(w)\overline{u}_{i}).
		\end{align*}
		Using Proposition \ref{prop:2} and Young's inequality, for $\forall X_{i}\in\Omega_{R}$,
		we obtain
		\begin{align}
		\dot{V}_{i}^{\zeta}\leq & -\frac{1}{2}\sum_{j=1}^{n-1}\zeta_{ij}^{2}+\zeta_{in}b_{in}(w)\overline{u}_{i}\nonumber \\
		& +\frac{\tilde{\gamma}_{i}}{2}||\overline{z}_{i}||^{2}+\frac{\tilde{\varrho}_{i}}{2}\sum_{j=1}^{n-1}||\tilde{\eta}_{ij}||^{2}+\mu_{i}^{*}(R)\zeta_{in}^{2}\label{eq:28}
		\end{align}
		where $\mu_{i}^{*}(R)$ is a positive constant related with $R$.
		
		\textit{5) Analysis of} $V_{i}$.
		
		Finally, based on (\ref{eq:27-1}) and (\ref{eq:28}), the derivative
		of $V_{i}$ can be computed as
		\begin{align*}
		\dot{V}_{i}\leq & -\frac{\tilde{\gamma}_{i}}{2}||\overline{z}_{i}||^{2}-\frac{\tilde{\varrho}_{i}}{2}\sum_{k=1}^{j}||\tilde{\eta}_{ij}||^{2}-\frac{1}{4}\sum_{j=1}^{n-1}\zeta_{ij}^{2}\\
		& -(b_{in}(w)K_{i}(R)-\mu_{i}^{*}(R))\zeta_{in}^{2}\\
		& +\zeta_{in}b_{in}(w)(\overline{u}_{i}-\overline{u}_{i}^{*})
		\end{align*}
		where $\overline{u}_{i}^{*}$ is given by (\ref{eq:26-2}).
		
		Hence, if 
		\[
		K_{i}(R)>\mu_{i}^{*}(R)/b_{in}(w)
		\]
		we obtain (\ref{eq:27-3}). The proof is completed.
	\end{IEEEproof}
	
	\subsection{Proof of Theorems \ref{thm:1-1-1} and \ref{thm:1-1-1-1}}
	\begin{IEEEproof}
		We only provide the proof for Theorem \ref{thm:1-1-1-1}. The proof
		of Theorem \ref{thm:1-1-1} follows. The proof is divided into the
		following steps.
		
		\textit{\textcolor{black}{Step 1).}}\textcolor{black}{{} }\textit{\textcolor{black}{Construction
				of the estimation error system.}}
		
		\textcolor{black}{Using the transformed system (\ref{eq:22-2}) and
			the observer (\ref{eq:27}), define the estimation error as $\tilde{\xi}_{ij}=\xi_{ij}-\hat{\xi}_{ij}(i=1,2,...,N;j=1,2,...,n)$.
			Then, the estimation error system is constructed as:}
		\begin{align*}
		\dot{\tilde{\xi}}_{i1}= & \tilde{\xi}_{i2}-\Gamma_{i}d_{1}\tilde{\xi}_{i1}+\Gamma_{i}d_{1}(\hat{e}_{i}(\overline{\tau}_{q}^{i})-e_{i}),\\
		\dot{\tilde{\xi}}_{i2}= & \tilde{\xi}_{i3}-\Gamma_{i}^{2}d_{2}\tilde{\xi}_{i1}+\Gamma_{i}^{2}d_{2}(\hat{e}_{i}(\overline{\tau}_{q}^{i})-e_{i}),\\
		\vdots\\
		\dot{\tilde{\xi}}_{in}= & -\Gamma_{i}^{n}d_{n}\tilde{\xi}_{i1}+\Gamma_{i}^{n}d_{n}(\hat{e}_{i}(\overline{\tau}_{q}^{i})-e_{i})\\
		& +\phi_{i}+(b_{in}(w)-\hat{b}_{in})\overline{u}_{i}.
		\end{align*}
		Let $\epsilon_{ij}=\Gamma_{i}^{n-j}\tilde{\xi}_{ij}(j=1,2,...,n)$.
		It follows that 
		\begin{equation}
		\dot{\epsilon}_{i}=\Gamma_{i}D_{\epsilon}\epsilon_{i}+H_{i}\label{eq:31-2}
		\end{equation}
		where $\epsilon_{i}=\mathrm{col}(\epsilon_{i1},\epsilon_{i2},...,\epsilon_{in})$,
		\[
		D_{\epsilon}=\left[\begin{array}{cccc}
		-d_{1} & 1\\
		\vdots &  & \ddots\\
		-d_{n-1} &  &  & 1\\
		-d_{n}
		\end{array}\right],
		\]
		\[
		H_{i}=H_{i1}+H_{i2}
		\]
		with 
		\[
		H_{i1}=\Gamma_{i}^{n}(\hat{e}_{i}(\overline{\tau}_{q}^{i})-e_{i}(t))\mathrm{col}(d_{1},...,d_{n}),
		\]
		\[
		H_{i2}=\mathrm{col}(0,0,...,\overline{\phi}_{i}+(b_{in}(w)-\hat{b}_{in})\overline{u}_{i}).
		\]
		
		\textit{\textcolor{black}{Step 2). Construction of the Lyapunov functions.}}
		
		\textcolor{black}{Note that since the design parameters $d_{1},...,d_{n}$
			are the coefficients of some Hurwitz polynomial $s^{n}+d_{1}s^{n-1}+\cdots+d_{n-1}s+d_{n}$,
			$D_{\epsilon}$ is Hurwtiz. This indicates that we can find a positive
			definite matrix $P$ such that $PD_{\epsilon}+D_{\epsilon}^{\mathrm{T}}P\leq-I$.
			Then, take the following Lyapunov function $V_{i}^{\epsilon}=\epsilon_{i}^{\mathrm{T}}P\epsilon_{i}$.
			The derivative of $V_{i}^{\epsilon}$ is given by }
		\begin{align}
		\dot{V}_{i}^{\epsilon}= & -\Gamma_{i}||\epsilon_{i}||^{2}+2\epsilon_{i}^{\mathrm{T}}PH_{i1}+2\epsilon_{i}^{\mathrm{T}}PH_{i2}.\label{eq:29-1}
		\end{align}
		For $\epsilon_{i}^{\mathrm{T}}PH_{i1}$, by Young's inequality and
		(\ref{eq:26-1}), we have 
		\begin{align}
		\epsilon_{i}^{\mathrm{T}}PH_{i1} & \leq\frac{\Gamma_{i}}{5}||\epsilon_{i}||^{2}+\sigma_{i1}(\Gamma_{i})(\hat{e}_{i}(\overline{\tau}_{q}^{i})-e_{i}(t))^{2}\label{eq:29}
		\end{align}
		where $\sigma_{i1}(\Gamma_{i})$ is a positive constant related with
		$\Gamma_{i}.$
		
		For $\epsilon_{i}^{T}PH_{i2}$, note that $\phi_{i}(0,...,0,\nu,w)=0$.
		Then when $X_{i}\in\Omega_{R}$ we have 
		\begin{align}
		\epsilon_{i}^{T}PH_{i2}\leq & \frac{\Gamma_{i}}{5}||\epsilon_{i}||^{2}+\sigma_{i2}(R)(\overline{z}_{i}^{2}+\sum_{j=1}^{n}(\tilde{\eta}_{ij}^{2}+\zeta_{ij}^{2}))\nonumber \\
		& +\sigma_{i2}(R)(\overline{u}_{i}-\overline{u}_{i}^{*})^{2}\label{eq:31}
		\end{align}
		where $\sigma_{i2}(R_{i})$ is a positive constant related with $R_{i}.$
		
		Finally, consider the following Lyapunov function in logarithm form
		\begin{equation}
		\mathcal{V}_{i}=V_{i}+\frac{\ln(1+V_{i}^{\epsilon})}{\ln(1+\varsigma_{i}(\Gamma_{i}))}.\label{eq:61}
		\end{equation}
		Assume $V_{i}^{\epsilon}(0)\leq R_{\epsilon}$ for $\forall X_{i}\in B_{r}$
		with a positive constant $R_{\epsilon}$. Then $\varsigma_{i}(\Gamma_{i})$
		is selected to be a polynomial function with respect to $\Gamma_{i}$
		such that $\frac{\ln(1+R_{\epsilon})}{\ln(1+\varsigma_{i}(\Gamma_{i}))}\leq\frac{\Delta_{R}}{2}$.
		
		Based on Lemma \ref{lem:For-system-(),} and (\ref{eq:29-1})-(\ref{eq:31}),
		the derivative of $\mathcal{V}_{i}$ is computed as:
		\begin{align}
		\mathcal{\dot{V}}_{i}\leq & -\tilde{\gamma}_{i}||\overline{z}_{i}||^{2}-\tilde{\varrho}_{i}\sum_{j=1}^{n}||\tilde{\eta}_{ij}||^{2}-\sum_{j=1}^{n}\zeta_{ij}^{2}\nonumber \\
		& +\Upsilon_{i1}+\Upsilon_{i2}+\Upsilon_{i3}+\Upsilon_{i4}\label{eq:36}
		\end{align}
		where
		\begin{align*}
		\Upsilon_{i1}= & -\frac{\Gamma_{i}}{5\ln(1+\varsigma_{i}(\Gamma_{i}))}\frac{||\mathit{\epsilon_{i}}||^{2}}{1+||P||||\mathit{\epsilon_{i}}||^{2}}\\
		& +\frac{2\sigma_{i2}(R)(\widetilde{u}_{i}-\overline{u}_{i}^{*})^{2}}{\ln(1+\varsigma_{i}(\Gamma_{i}))},
		\end{align*}
		\begin{align*}
		\Upsilon_{i2}= & \frac{2\sigma_{i2}(R)\left(\overline{z}_{i}^{2}+\sum_{j=1}^{2}(\tilde{\eta}_{ij}^{2}+\zeta_{ij}^{2})\right)}{\ln(1+\varsigma_{i}(\Gamma_{i}))},
		\end{align*}
		\begin{align*}
		\Upsilon_{i3} & =\frac{2\sigma_{i2}(R)(\overline{u}_{i}-\widetilde{u}_{i})^{2}}{\ln(1+\varsigma_{i}(\Gamma_{i}))},
		\end{align*}
		\begin{align*}
		\Upsilon_{i4}= & \frac{2\sigma_{i1}(\Gamma_{i})(\hat{e}_{i}(\overline{\tau}_{q}^{i})-e_{i}(t))^{2}}{\ln(1+\varsigma_{i}(\Gamma_{i}))},
		\end{align*}
		\begin{align*}
		\widetilde{u}_{i} & =\omega_{i}(t)-\Psi_{in}\eta_{in}(t)=\mathrm{sat}_{\mathcal{R}}(K_{i}(R)\hat{\zeta}_{in}(t)),
		\end{align*}
		\[
		\overline{u}_{i}=u_{i}(t)-\Psi_{in}\eta_{in}(t).
		\]
		
		From (\ref{eq:26-2}), (\ref{eq:29-2-1}) and Lemma \ref{lem:1)-(Gronwall's-inequality)},
		we know there exists a positive constant $\sigma_{i3}(R)$ and sufficient
		large $\mathcal{R}$ such that
		\begin{align*}
		(\widetilde{u}_{i}-\overline{u}_{i}^{*})^{2}\leq & \sigma_{i3}(R)\mathrm{min}\{||\mathit{\epsilon_{i}}||^{2},1\}\leq\frac{\sigma_{i3}(R)||\mathit{\epsilon_{i}}||^{2}}{1+||P||||\mathit{\epsilon_{i}}||^{2}}.
		\end{align*}
		\textcolor{black}{Using this for (\ref{eq:36}) and noting that $\varsigma_{i}(\Gamma_{i})$
			is a polynomial function with respect to $\Gamma_{i}$,} there exists
		a sufficiently large $\Gamma_{i}$ such that
		\begin{align}
		\mathcal{\dot{V}}_{i}\leq & -\frac{\tilde{\gamma}_{i}}{2}||\overline{z}_{i}||^{2}-\frac{\tilde{\varrho}_{i}}{2}\sum_{k=1}^{n}||\tilde{\eta}_{ij}||^{2}-\frac{1}{2}\sum_{j=1}^{n}\zeta_{ij}^{2}-\sigma_{i4}||\mathit{\epsilon_{i}}||^{2}\nonumber \\
		& +\Upsilon_{i3}+\Upsilon_{i4}\label{eq:38-1}
		\end{align}
		where $\sigma_{i4}$ is a positive constant.
		
		\textit{\textcolor{black}{Step 3). Taking the PET mechanism into consideration}}
		
		We will have an analysis on the terms $\Upsilon_{i3},\Upsilon_{i4}$
		in (\ref{eq:38-1}) by taking the PET mechanism into consideration.
		In the following, we suppose $t\in[\tau_{p}^{i},\tau_{p+1}^{i})$.
		
		Using (\ref{eq:26-4-2})-(\ref{eq:32}), $\overline{u}_{i}-\widetilde{u}_{i}$
		in $\Upsilon_{i3}$ is computed as:
		\[
		\overline{u}_{i}-\widetilde{u}_{i}=\omega_{i}(\overline{\varsigma}_{m}^{i})-\omega_{i}(t)=\omega_{i}(\overline{\varsigma}_{m}^{i})-\omega_{i}(\tau_{p}^{i})+\omega_{i}(\tau_{p}^{i})-\omega_{i}(t)
		\]
		\textcolor{black}{where $\overline{\varsigma}_{m}^{i}$ denotes the
			latest event-triggered time instant for the data transmission between
			the controller and the plant.}
		
		By the event triggered condition (\ref{eq:26-1-1-1}), we have
		\begin{align}
		|\overline{u}_{i}-\widetilde{u}_{i}| & \leq\iota_{\omega}|\omega_{i}(\tau_{p}^{i})|+|\omega_{i}(\tau_{p}^{i})-\omega_{i}(t)|\nonumber \\
		& \leq(1+\iota_{\omega})|\omega_{i}(\tau_{p}^{i})-\omega_{i}(t)|+\iota_{\omega}|\omega_{i}(t)|.\label{eq:65}
		\end{align}
		Meanwhile, by (\ref{eq:26-1}) and Theorem \ref{thm:1-1}, $\hat{e}_{i}(\overline{\tau}_{q}^{i})-e_{i}(t)$
		in $\Upsilon_{i4}$ is computed as:
		\begin{align}
		& |\hat{e}_{i}(\overline{\tau}_{q}^{i})-e_{i}(t)|\nonumber \\
		\leq & |\hat{e}_{i}(\overline{\tau}_{q}^{i})-\hat{e}_{i}(\tau_{p}^{i})|+|\hat{e}_{i}(\tau_{p}^{i})-e_{i}(\tau_{p}^{i})|+|e_{i}(\tau_{p}^{i})-e_{i}(t)|\nonumber \\
		\leq & \iota_{e}|\hat{e}_{i}(\tau_{p}^{i})|+\mathrm{e}^{-\gamma_{\nu}\tau_{p}^{i}}+|e_{i}(\tau_{p}^{i})-e_{i}(t)|,\label{eq:64}
		\end{align}
		\begin{align}
		|\hat{e}_{i}(\tau_{p}^{i})| & \leq|\hat{e}_{i}(\tau_{p}^{i})-e_{i}(\tau_{p}^{i})|+|e_{i}(\tau_{p}^{i})-e_{i}(t)|+|e_{i}(t)|\nonumber \\
		& \leq\mathrm{e}^{-\gamma_{\nu}\tau_{p}^{i}}+|e_{i}(\tau_{p}^{i})-e_{i}(t)|+|e_{i}(t)|\label{eq:65-1}
		\end{align}
		\textcolor{black}{where $\overline{\tau}_{q}^{i}$ denotes the latest
			event-triggered time instant for the data transmission between the
			sensor and the plant.}
		
		Using (\ref{eq:65})-(\ref{eq:65-1}) for $\Upsilon_{i3},\Upsilon_{i4}$
		in (\ref{eq:38-1}), we have
		\begin{align}
		\mathcal{\dot{V}}_{i}\leq & -\frac{\tilde{\gamma}_{i}}{2}||\overline{z}_{i}||^{2}-\frac{\tilde{\varrho}_{i}}{2}\sum_{k=1}^{2}||\tilde{\eta}_{ij}||^{2}-\frac{1}{2}\sum_{j=1}^{n}\zeta_{ij}^{2}-\sigma_{i4}||\mathit{\epsilon_{i}}||^{2}\nonumber \\
		& +\sigma_{i5}(R)\left((\omega_{i}(\tau_{p}^{i})-\omega_{i}(t))^{2}+\iota_{\omega}^{2}\omega_{i}^{2}(t)\right)\nonumber \\
		& +\sigma_{i6}(\Gamma_{i})\left((e_{i}(\tau_{p}^{i})-e_{i}(t))^{2}+\iota_{e}^{2}e_{i}^{2}(t)+\mathrm{e}^{-2\gamma_{\nu}t}\right)\label{eq:78}
		\end{align}
		where $\sigma_{i5}(R),\sigma_{i6}(\Gamma_{i})$ are positive constants
		related with $R,\Gamma_{i}.$
		
		\textit{\textcolor{black}{Step 4). Convergence analysis}}
		
		\textcolor{black}{Let $\mathcal{X}_{i}=\mathrm{col}(\overline{z}_{i},\tilde{\eta}_{i1},\tilde{\eta}_{in},\zeta_{i1},...,\zeta_{in},\mathit{\epsilon_{i}})$.
			From (\ref{eq:22-1}) and (\ref{eq:31-2}), we have}
		\begin{align}
		\mathcal{\dot{X}}_{i}= & \mathcal{A}_{i}\mathcal{X}_{i}+\mathcal{B}_{i}(\omega_{i}(\overline{\varsigma}_{m}^{i})-\omega_{i}(\tau_{p}^{i}))+\mathcal{B}_{i}(\omega_{i}(\tau_{p}^{i})-\omega_{i}(t))\nonumber \\
		& +\mathcal{C}_{i}(\hat{e}_{i}(\overline{\tau}_{q}^{i})-\hat{e}_{i}(\tau_{p}^{i}))+\mathcal{C}_{i}(\hat{e}_{i}(\tau_{p}^{i})-e_{i}(\tau_{p}^{i}))\nonumber \\
		& +\mathcal{C}_{i}(e_{i}(\tau_{p}^{i})-e_{i}(t))\nonumber \\
		& +\mathcal{D}_{i}(\mathcal{X}_{i})\label{eq:40-3}
		\end{align}
		where $\mathcal{A}_{i},\mathcal{B}_{i},\mathcal{C}_{i}$ are constant
		matrices, $\mathcal{D}_{i}(\mathcal{X}_{i})$ is a nonlinear function
		with respect to $\mathcal{X}_{i}$ such that
		\begin{align}
		||\mathcal{D}_{i}(\mathcal{X}_{i})|| & \leq c_{1}||\mathcal{X}_{i}||\label{eq:41}
		\end{align}
		for bounded $\mathcal{X}_{i}$ where $c_{1}>0$ is a positive constant.
		
		\textcolor{black}{For $\omega_{i}(\overline{\varsigma}_{m}^{i})-\omega_{i}(\tau_{p}^{i})$
			and $\omega_{i}(\tau_{p}^{i})-\omega_{i}(t)$ in (\ref{eq:40-3}),
			using (\ref{eq:26-1-1-1}), (\ref{eq:32}) and (\ref{eq:6-1}), we
			have}
		\begin{align}
		& |\omega_{i}(\overline{\varsigma}_{m}^{i})-\omega_{i}(\tau_{p}^{i})|\leq\iota_{\omega}|\omega_{i}(\tau_{p}^{i})|,\label{eq:70}
		\end{align}
		\begin{align}
		|\omega_{i}(\tau_{p}^{i})|= & |\mathrm{sat}_{\mathcal{R}}(K_{i}(R)\hat{\zeta}_{in}(\tau_{p}^{i}))\nonumber \\
		& +\Psi_{in}(\tilde{\eta}_{ij}(\tau_{p}^{i})+\theta_{ij}(\nu(\tau_{p}^{i}),w)+b^{-1}N_{ij}\overline{x}_{ij}(\tau_{p}^{i}))|\nonumber \\
		\leq & c_{2}||\mathcal{X}_{i}(\tau_{p}^{i})||+c_{3},\label{eq:71}
		\end{align}
		\begin{align}
		& |\omega_{i}(\tau_{p}^{i})-\omega_{i}(t)|\nonumber \\
		= & |\mathrm{sat}_{\mathcal{R}}(K_{i}(R)\hat{\zeta}_{in}(\tau_{p}^{i}))+\Psi_{in}\eta_{in}(\tau_{p}^{i})\nonumber \\
		& -(\mathrm{sat}_{\mathcal{R}}(K_{i}(R)\hat{\zeta}_{in})+\Psi_{in}\eta_{in})|\nonumber \\
		\leq & c_{4}||\mathcal{X}_{i}-\mathcal{X}_{i}(\tau_{p}^{i})||+|\Psi_{in}(\tilde{\eta}_{ij}+\theta_{ij}(\nu,w)+b^{-1}N_{ij}\overline{x}_{ij})\nonumber \\
		& -\Psi_{in}(\tilde{\eta}_{ij}(\tau_{p}^{i})+\theta_{ij}(\nu(\tau_{p}^{i}),w)+b^{-1}N_{ij}\overline{x}_{ij}(\tau_{p}^{i}))|\nonumber \\
		\leq & c_{5}||\mathcal{X}_{i}-\mathcal{X}_{i}(\tau_{p}^{i})||+\delta_{i1}(\mathcal{T}^{i})\label{eq:72}
		\end{align}
		where $c_{2},c_{3},c_{4},c_{5}$ are positive constants, $\delta_{i1}(\mathcal{T}^{i})$
		is an increasing function with $\delta_{i1}(0)=0$. 
		
		\textcolor{black}{For $\hat{e}_{i}(\overline{\tau}_{q}^{i})-\hat{e}_{i}(\tau_{p}^{i})$,
			$\hat{e}_{i}(\tau_{p}^{i})-e_{i}(\tau_{p}^{i})$ and $e_{i}(\tau_{p}^{i})-e_{i}(t)$
			in (\ref{eq:40-3}),} from (\ref{eq:26-1}) and Theorem \ref{thm:1},
		noting that $\zeta_{i1}=\xi_{i1}=e_{i}$, we have
		\begin{align}
		& |(\hat{e}_{i}(\overline{\tau}_{q}^{i})-\hat{e}_{i}(\tau_{p}^{i}))+(\hat{e}_{i}(\tau_{p}^{i})-e_{i}(\tau_{p}^{i}))|\nonumber \\
		\leq & c_{6}\iota_{e}|\hat{e}_{i}(\tau_{p}^{i})|+c_{6}\mathrm{e}^{-\gamma_{\nu}t}\nonumber \\
		\leq & c_{6}\iota_{e}|\hat{e}_{i}(\tau_{p}^{i})-e_{i}(\tau_{p}^{i})|+c_{6}\iota_{e}|e_{i}(\tau_{p}^{i})|+c_{6}\mathrm{e}^{-\gamma_{\nu}t}\nonumber \\
		\leq & c_{7}\iota_{e}||\mathcal{X}_{i}(\tau_{p}^{i})||+c_{7}\mathrm{e}^{-\gamma_{\nu}t},\label{eq:54}
		\end{align}
		\begin{equation}
		|e_{i}(\tau_{p}^{i})-e_{i}(t)|\leq c_{8}||\mathcal{X}_{i}(t)-\mathcal{X}_{i}(\tau_{p}^{i})||\label{eq:42}
		\end{equation}
		where $c_{6},c_{7},c_{8}>0$ are positive constants. 
		
		Then, integrating (\ref{eq:40-3}) on time interval $[\tau_{p}^{i},\tau_{p+1}^{i})$
		and using (\ref{eq:70})-(\ref{eq:42}), we have
		\begin{align*}
		& ||\mathcal{X}_{i}(t)-\mathcal{X}_{i}(\tau_{p}^{i})||\\
		\leq & \int_{\tau_{p}^{i}}^{t}c_{9}||\mathcal{X}_{i}(\tau)-\mathcal{X}_{i}(\tau_{p}^{i})||d\tau+\mathcal{T}^{i}c_{10}||\mathcal{X}_{i}(\tau_{p}^{i})||\\
		& +c_{11}\mathrm{e}^{-\gamma_{\nu}\tau_{p}^{i}}+\delta_{i2}(\mathcal{T}^{i})
		\end{align*}
		where $c_{9},c_{10},c_{11}>0$ are positive constants and $\delta_{i2}(\mathcal{T}^{i})$
		is an increasing function with $\delta_{i2}(0)=0$. 
		
		Using Gronwall\textquoteright s inequality, we have
		\begin{align*}
		& ||\mathcal{X}_{i}(t)-\mathcal{X}_{i}(\tau_{p}^{i})||\\
		\leq & (\mathcal{T}^{i}c_{10}||\mathcal{X}_{i}(\tau_{p}^{i})||+c_{11}\mathrm{e}^{-\gamma_{\nu}\tau_{p}^{i}}+\delta_{i2}(\mathcal{T}^{i}))\mathrm{e}^{-c_{9}\mathcal{T}^{i}}.
		\end{align*}
		It follows that
		\begin{align*}
		& ||\mathcal{X}_{i}(t)-\mathcal{X}_{i}(\tau_{p}^{i})||\\
		\leq & \mathcal{T}^{i}c_{10}\mathrm{e}^{-c_{9}\mathcal{T}^{i}}||\mathcal{X}_{i}(t)-\mathcal{X}_{i}(\tau_{p}^{i})||\\
		& +\mathcal{T}^{i}c_{10}\mathrm{e}^{-c_{9}\mathcal{T}^{i}}||\mathcal{X}_{i}(t)||+c_{11}\mathrm{e}^{-\gamma_{\nu}\tau_{p}^{i}}\mathrm{e}^{-c_{9}\mathcal{T}^{i}}+\delta_{i2}(\mathcal{T}^{i})\mathrm{e}^{-c_{9}\mathcal{T}^{i}}.
		\end{align*}
		Then, when $\mathcal{T}^{i}$ is small enough, we have
		\begin{align*}
		& ||\mathcal{X}_{i}(t)-\mathcal{X}_{i}(\tau_{p}^{i})||\\
		\leq & \Xi_{i}(\mathcal{T}^{i})||\mathcal{X}_{i}(t)||+c_{12}\mathrm{e}^{-\gamma_{\nu}\tau_{p}^{i}}+\delta_{i3}(\mathcal{T}^{i})
		\end{align*}
		where $\Xi_{i}(\mathcal{T}^{i})$ is an increasing function with $\Xi_{i}(0)=0$,
		$c_{12}$ is a positive constant and $\delta_{i3}(\mathcal{T}^{i})$
		is an increasing function with $\delta_{i3}(0)=0$. 
		
		\textcolor{black}{Next, using the above inequality for $\omega_{i}(\tau_{p}^{i})-\omega_{i}(t)$
			and $e_{i}(\tau_{p}^{i})-e_{i}(t)$ in (\ref{eq:78}), we can conclude
			that there exists a sufficiently small sampling period $\mathcal{T}^{i}$
			and $\iota_{e},\iota_{\omega}$ such that }
		\begin{align}
		\mathcal{\dot{V}}_{i}\leq & -\frac{\tilde{\gamma}_{i}}{4}||\overline{z}_{i}||^{2}-\frac{\tilde{\varrho}_{i}}{4}\sum_{k=1}^{n}||\tilde{\eta}_{ij}||^{2}-\frac{1}{4}\sum_{j=1}^{n}\zeta_{ij}^{2}-\frac{\sigma_{i6}}{2}||\mathit{\epsilon_{i}}||^{2}\nonumber \\
		& +c_{13}\mathrm{e}^{-2\gamma_{\nu}t}+\delta_{i4}(\iota_{e},\iota_{\omega},\mathcal{T}^{i})\nonumber \\
		\leq & -c_{14}\mathcal{V}_{i}+c_{13}\mathrm{e}^{-2\gamma_{\nu}t}+\delta_{i4}(\iota_{e},\iota_{\omega},\mathcal{T}^{i})
		\end{align}
		where $c_{13},c_{14}>0$ are positive constants and $\delta_{i4}(\iota_{e},\iota_{\omega},\mathcal{T}^{i})$
		is an increasing function with $\delta_{i4}(0,0,0)=0$. 
		
		By solving the above equation, we have
		\[
		\mathcal{V}_{i}(t)\leq\mathcal{V}_{i}(0)+c_{15}\left(\frac{1}{2\gamma_{\nu}}+\delta_{i4}(\iota_{e},\iota_{\omega},\mathcal{T}^{i})\right)
		\]
		where $c_{15}$ is a positive constant. This means that there exists
		a sufficient large $\gamma_{\nu}$ and small $\iota_{e},\iota_{\omega},\mathcal{T}^{i}$
		such that $\mathcal{V}_{i}(t)\leq\overline{R}+\Delta_{R}=R$. Therefore,
		$X_{i}$ will always remain in the set $\Omega_{R}$. Meanwhile, $\mathcal{V}_{i}(t)$
		will converge to the set $\delta_{i4}(\iota_{e},\iota_{\omega},\mathcal{T}^{i})/c_{14}$
		exponentially. This completes the proof.
	\end{IEEEproof}

\end{document}